\documentclass[lettersize,journal]{IEEEtran}

\usepackage{amsmath,amsfonts}
\usepackage{algorithmic}
\usepackage{algorithm}
\usepackage{mathtools}
\usepackage{array}
\usepackage{siunitx}
\usepackage{amsmath}
\usepackage{amssymb}
\usepackage[caption=false,font=normalsize,labelfont=sf,textfont=sf]{subfig}
\usepackage{textcomp}
\usepackage{url}
\usepackage{verbatim}
\usepackage{graphicx}
\usepackage{epstopdf}
\usepackage{multirow}
\usepackage{epstopdf}
\usepackage{caption}
\usepackage{lipsum}
\usepackage[utf8]{inputenc}
\usepackage{caption} 
\usepackage{authblk}
\usepackage{bm}
\usepackage[font=small]{caption}
\usepackage{xcolor}
\usepackage{hyperref}
\usepackage{makecell}
\usepackage{booktabs}
\usepackage{cite}

\DeclareGraphicsExtensions{.jpg,.png,.pdf}
\begin{document}
\title{Energy Efficiency Maximization in RSMA-Based LEO Satellite Networks Assisted by UAV-Mounted BD-Active RIS: A Hybrid DRL Strategy}

\author{Rahman Saadat Yeganeh$^1$, Hamid Behroozi$^{1}$, Mohammad Javad Omidi$^{2,3}$, Mohammad Robat Mili$^{4}$, Symeon Chatzinotas$^{5},$ \IEEEmembership{Fellow, IEEE}, and Eduard A. Jorswieck$^{6},$ \IEEEmembership{Fellow, IEEE}

\thanks{$^1$Department of Electrical Engineering, Sharif University of Technology, Tehran, Iran
(Emails: {rahman.saadat, behroozi}@sharif.edu)}

\thanks{$^2$Department of Electronics and Communication Engineering, Kuwait College of Science and Technology, Doha, Kuwait
(Email: m.omidi@kcst.edu.kw).}

\thanks{$^3$Department of Electrical and Computer Engineering, Isfahan University of Technology, Isfahan, Iran
(Email: omidi@iut.ac.ir).}

\thanks{$^4$Department of Computer and Engineering, Pasargad Institute for Advanced Innovative Solutions (PIAIS), Tehran, Iran
(Email: mohammad.robatmili@gmail.com).}

\thanks{$^5$Interdisciplinary Centre for Security, Reliability and Trust, University of Luxembourg
(Email: Symeon.Chatzinotas@uni.lu).}

\thanks{$^6$Institute for Communications Technology,
Technische Universität Braunschweig,
Braunschweig, Germany
(Email: e.jorswieck@tu-braunschweig.de)}

}

\maketitle

\begin{abstract}
This paper proposes a novel non-terrestrial communication architecture that integrates Rate-Splitting Multiple Access (RSMA) with a UAV-mounted Beyond-Diagonal Active Reconfigurable Intelligent Surface (BD-ARIS) under the coverage of a Low Earth Orbit (LEO) satellite. The BD-ARIS employs a group-connected structure to provide signal amplification and enhanced beamforming flexibility, while RSMA splits user messages into common and private parts to improve SE and interference management. A joint optimization of satellite beamforming, UAV placement, power allocation, and RIS configuration is formulated to maximize the system's overall energy efficiency. To solve the resulting non-convex, high-dimensional problem, we first develop a model-based iterative algorithm using Dinkelbach's transform, weighted minimum mean-square error reformulation, successive convex approximation, and sequential quadratic programming. We then propose two deep reinforcement learning (DRL) approaches: a standard trust-region policy optimization (TRPO) baseline and a novel Dinkelbach-Guided TRPO (DG-TRPO) scheme. DG-TRPO is initialized via supervised pretraining on Dinkelbach solutions and subsequently fine-tuned with TRPO in noisy, time-varying environments. \textcolor{black}{Simulation results show that DG-TRPO closely matches the model-based benchmark while providing sub-millisecond inference, making it suitable for real-time LEO operation.} It also exhibits superior robustness to CSI uncertainty and satellite mobility, and the RSMA BD-ARIS framework significantly outperforms benchmarks employing NOMA or conventional RIS designs. These results establish the proposed architecture and DG-TRPO algorithm as a scalable, energy-efficient solution for future 6G massive IoT non-terrestrial networks.
\end{abstract}


\section{Introduction}

\subsection{Background}

The exponential growth of Internet of Things (IoT) applications across diverse sectors such as smart cities, healthcare, and industrial automation has driven the urgent need for wireless networks that offer seamless, energy-efficient, and globally accessible connectivity. Sixth-generation (6G) networks are envisioned to meet these ambitious requirements through the integration of cutting-edge technologies, including Non-Terrestrial Networks (NTNs), LEO satellite constellations, RISs, and Unmanned Aerial Vehicles (UAVs) \cite{8879484,you2021towards}.

Among these technologies, LEO satellites play a central role by providing low-latency, wide-area coverage essential for massive IoT deployment. However, their high orbital speeds result in short contact times with Ground Terminals, particularly in remote or obstructed areas with sparse terrestrial infrastructure \cite{cao2024toward}. To address these challenges, RISs have emerged as a promising solution capable of dynamically reconfiguring the wireless propagation environment without the need for expensive hardware upgrades \cite{xu2023active}.

Building upon the conventional RIS concept, BD-ARIS has been introduced to overcome the limitations of passive surfaces. BD-ARIS enhances signal coverage and link reliability by amplifying incident signals and enabling flexible beamforming via group-connected architectures \cite{khan2025surveydiagonalrisenabled}. Mounting BD-ARIS units on UAV platforms further extends system adaptability, enabling dynamic and location-aware relay operations that are crucial for maintaining reliable satellite-to-ground communications in rapidly changing and complex environments.

At the physical layer, RSMA has gained considerable attention as an effective transmission strategy. By splitting user messages into common and private parts, RSMA provides a robust framework to manage user interference and enhance both SE and fairness, even under heterogeneous channel conditions \cite{9145189}. Nevertheless, incorporating RSMA into UAV-assisted, BD-ARIS-enhanced satellite systems introduces significant design complexities, including high-dimensional, strongly coupled, and non-convex optimization problems across spatial, power, and rate dimensions.

To tackle these challenges, we first develop a model‑based optimization algorithm that combines Dinkelbach's transform with weighted minimum mean‑square error (WMMSE) reformulation, successive convex approximation (SCA), and sequential quadratic programming (SQP) to obtain high-quality stationary solutions. While effective, its high computational delay limits its applicability in dynamic LEO scenarios where the channel coherence time is only a few milliseconds. We therefore introduce a hybrid DRL strategy, Dinkelbach‑Guided TRPO (DG‑TRPO), in which an actor network is first pretrained on the outputs of the model‑based solver and then fine‑tuned with a standard TRPO procedure. This approach retains the accuracy of model‑based optimization while gaining the real‑time inference speed (sub‑millisecond) and robustness to CSI imperfections that are essential for practical LEO satellite communications.

Overall, the integration of LEO satellite networks for wide coverage, UAV-mounted BD-ARIS for intelligent and flexible signal reflection and amplification, RSMA for enhanced multiple access, and an optimization‑driven hybrid DRL framework forms a holistic and powerful architecture. This synergy paves the way for the development of sustainable, adaptive, and high‑capacity 6G communication infrastructures, capable of supporting the next generation of global IoT applications and beyond.
\footnote{A preliminary version of this work was made available in~\cite{yeganeh2025energy}.}

\subsection{Related Works}

The integration of satellite communication with intelligent surfaces and UAV platforms has gained significant research attention in recent years. Various studies have explored the potential of LEO satellites to enhance global connectivity \cite{RSY2025}. In~\cite{cao2024toward}, the authors provide a comprehensive review of LEO satellite systems, highlighting their benefits in terms of latency and coverage, alongside challenges such as intermittent visibility and rapid handover requirements in terrestrial environments. \textcolor{black}{Recent Direct-to-Smartphone studies further emphasize high-dynamic air-interface adaptation and large-array multi-beam management as key challenges for LEO-based 6G NTN systems~\cite{he2024direct}.}

To mitigate coverage gaps and signal attenuation in satellite-to-ground communications, RISs have been proposed as auxiliary infrastructure. Passive RISs have been extensively studied in terrestrial scenarios due to their EE and low cost~\cite{huang2019reconfigurable}. However, their limited ability to manipulate weak long-distance signals, such as those from satellites, has motivated the development of active RISs~\cite{zhang2022active}, which incorporate amplification circuits to enhance signal strength.

Recently, the integration of RISs into satellite communication systems has attracted notable interest. Dong~et~al.~\cite{9500640} jointly optimized the transmit beamforming and RIS phase shifts to maximize the weighted sum rate in integrated satellite-terrestrial networks. In~\cite{10008708}, RIS orientation and passive beamforming were optimized to enhance average throughput in LEO systems. Lee~et~al.~\cite{9621010} maximized the received SNR in RIS-aided LEO satellite communications, and~\cite{9849035} improved channel gains via joint transmit and receive beamforming designs. \textcolor{black}{Security-oriented RIS designs have also been studied for satellite-terrestrial networks, including self-powered absorptive RISs under secrecy-rate and harvested-power constraints~\cite{zhi2024self}.}

The optimization of LEO satellite communication networks has also been extensively studied, including cache-enabled throughput maximization~\cite{9678021}, fairness and reliability enhancement via dual decomposition~\cite{8951285}, cooperative terrestrial-satellite rate maximization~\cite{9583591}, gateway placement via particle swarm optimization~\cite{9380214}, resource allocation in integrated networks~\cite{9080558}, and learning-based efficiency improvements~\cite{9420293}. \textcolor{black}{Recent learning-driven NTN studies have further considered multi-RIS-assisted OTFS resource scheduling through actor-critic DRL for secure SAGINs~\cite{feng2026towards}.}

In parallel, the integration of non-orthogonal multiple access (NOMA) techniques into LEO satellite networks has been explored to improve SE. A DRL-driven resource optimization approach for effective capacity maximization in NOMA-based LEO systems was proposed in~\cite{9838335}.

\textcolor{black}{In power-domain NOMA, users typically rely on different power levels and successive interference cancellation (SIC) according to a prescribed decoding order, whereas one-layer RSMA splits each message into common and private parts, providing greater flexibility in interference management.}

Focusing on IoT networks, the commensal symbiotic radio (CSR) system was introduced in~\cite{10138619} to enhance energy efficiency by enabling passive symbiotic backscatter devices. Moreover, an active STAR-RIS was employed in~\cite{10879566} for CSR-aided 6G networks, where a DRL approach using PPO, TD3, and A3C jointly designed beamforming and scheduling, significantly enhancing network throughput. \textcolor{black}{RSMA has also been investigated for massive IoT access in satellite-aerial integrated networks, where joint beamforming and resource optimization are performed under SINR and power constraints~\cite{lin2021supporting}.}

In the domain of RSMA and reconfigurable surfaces, the work in~\cite{10288244} proposed a general optimization framework for RSMA in BD-RIS-assisted URLLC systems, demonstrating significant gains under system overload and short packet transmissions. Furthermore,~\cite{10411856} explored the synergy between RSMA and BD-RIS to improve coverage and reduce antenna requirements, showing that BD-RIS-aided RSMA outperforms conventional SDMA schemes under imperfect CSI.

\textcolor{black}{More closely related to our work,~\cite{wu2025ris} combines RSMA, RIS, UAV-assisted SATINs, and LSTM-PPO for secrecy energy-efficiency optimization, while~\cite{guo2024joint} employs federated DRL for joint beamforming, RIS, UAV trajectory, and power-splitting optimization. In contrast, we consider a UAV-mounted group-connected BD-ARIS and Dinkelbach-guided TRPO with model-based pretraining for energy-efficiency optimization under LEO mobility and imperfect CSI.}

\subsection{Contributions}

Motivated by the limitations of traditional RIS-assisted non-terrestrial networks (NTNs) and the growing need for intelligent, adaptive multiple access in dynamic 6G environments, this paper introduces a holistic framework that combines BD‑ARIS, RSMA, model‑based optimization, and a novel hybrid DRL method for energy‑efficient satellite communication. The main contributions are:

\begin{itemize}
    \item \textbf{RSMA architecture for UAV-mounted BD-ARIS-assisted LEO downlink:} We design the first RSMA‑enabled UAV‑mounted group‑connected BD‑ARIS relaying system for LEO satellite downlinks. This architecture jointly exploits RSMA interference management and beyond‑diagonal active beamforming to boost energy efficiency.
\textcolor{black}{\item \textbf{Tractable model-based optimization algorithm:} We propose a convergent iterative solver based on Dinkelbach's fractional programming, WMMSE rate reformulation, SCA, and SQP, which provides high-quality stationary solutions for the non-convex EE maximization problem and serves as an offline model-based benchmark.}
\textcolor{black}{\item \textbf{Dinkelbach-Guided TRPO (DG-TRPO) agent:} We introduce a hybrid learning scheme where the actor is pretrained with Dinkelbach-AO outputs and then fine-tuned via TRPO. DG-TRPO closely matches the model-based benchmark while offering sub-millisecond inference, enabling real-time operation under satellite mobility and CSI errors.}
\item \textbf{Robustness and scalability validation:} Extensive simulations under varying powers, distances, CSI errors, user densities, and RIS sizes demonstrate that the RSMA BD‑ARIS combination consistently outperforms NOMA and conventional RIS, and that DG‑TRPO scales gracefully with system size.
    \item \textbf{Design insights for 6G NTN:} We distill practical recommendations for integrating RSMA and BD‑ARIS in future LEO IoT networks, emphasizing the benefits of combining physical‑layer reconfigurability with offline‑trained, fast‑inference DRL agents.
\end{itemize}

This paper is structured as follows. Section~II presents the system model for the BD‑ARIS‑assisted satellite communication network. Section~III formulates the energy efficiency maximization problem. Section~IV develops the model‑based optimization algorithm via Dinkelbach, WMMSE, SCA, and SQP. Section~V describes the Markov decision process framework and the two deep reinforcement learning approaches: standard TRPO and the proposed Dinkelbach‑Guided TRPO. Section~VI provides comprehensive simulation results and performance comparisons. Finally, Section~VII concludes the paper.

\begin{table}
    \centering
    \caption{List of abbreviations.}
    \begin{tabular}{|c|c|}
        \hline
        ARIS & Active RIS \\
        \hline
        BD-ARIS & Beyond Diagonal Active RIS \\
        \hline
        CSCG & Circularly Symmetric Complex Gaussian \\
        \hline
        CSI & Channel State Information \\
        \hline
        DG-TRPO & Dinkelbach-Guided Trust Region Policy Optimization \\
        \hline
        DRL & Deep Reinforcement Learning \\
        \hline
        EE & Energy Efficiency \\
        \hline
        IoT & Internet of Things \\
        \hline
        LoS & Line of Sight \\
        \hline
        LEO & Low Earth Orbit \\
        \hline
        MIMO & Multiple Input Multiple Output \\
        \hline
        NOMA & Non-Orthogonal Multiple Access \\
        \hline
        NTN & Non-Terrestrial Networks \\
        \hline
        QoS & Quality of Service \\
        \hline
        RIS & Reconfigurable Intelligent Surface \\
        \hline
        RSMA & Rate Splitting Multiple Access \\
        \hline
        SIC & Successive Interference Cancellation \\
        \hline
        SE & Spectral Efficiency \\
	\hline
        TRPO & Trust Region Policy Optimization \\
                \hline
        WMMSE  & Weighted Minimum Mean-Square Error \\
        \hline
    \end{tabular}
\end{table}

\section{System Model}

\begin{figure}[!t]
\centering
\includegraphics[width=.9\columnwidth]{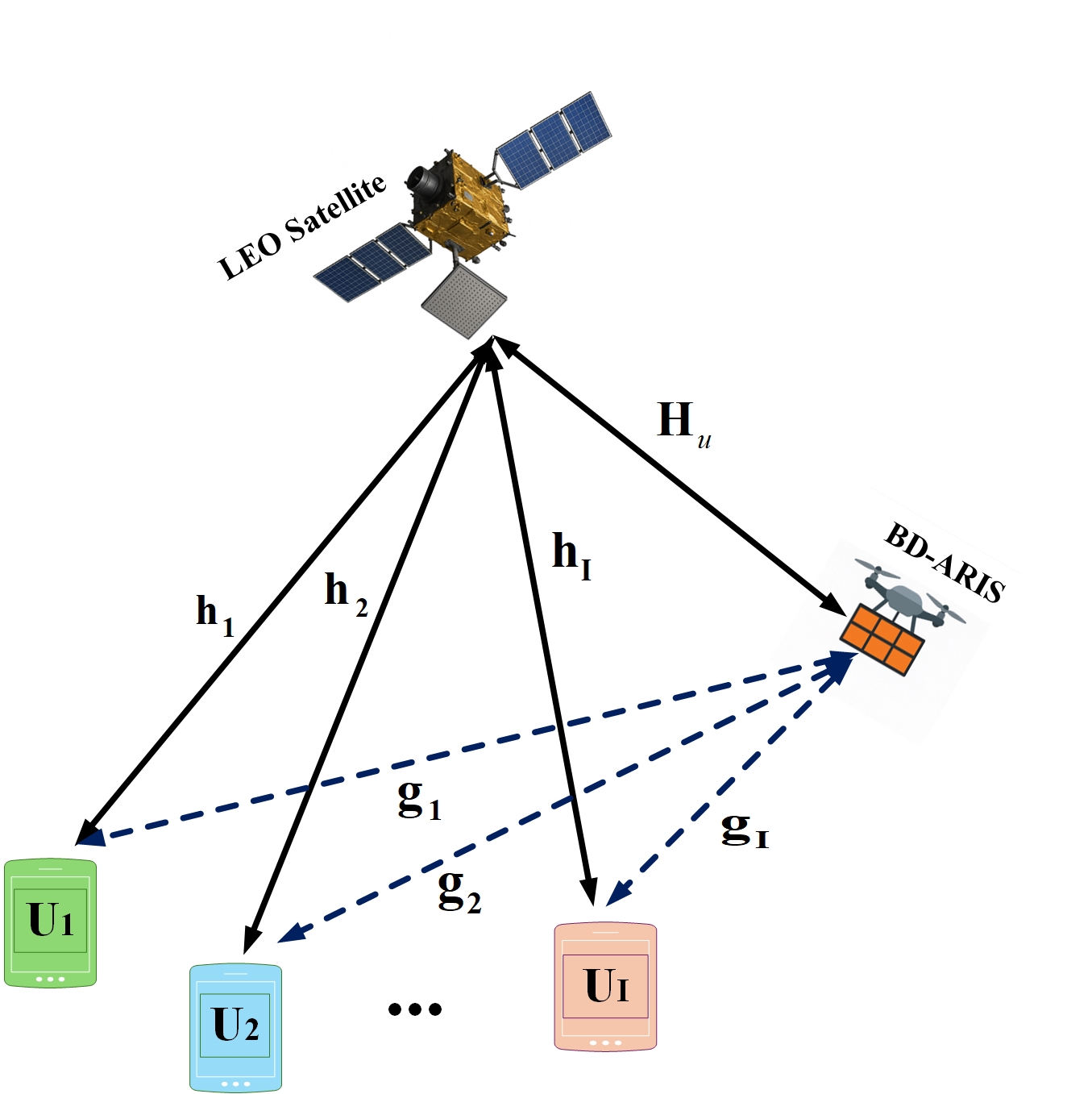}
\caption{System model of a UAV-mounted BD-ARIS-assisted LEO satellite communication network with multiple ground users.}
\label{fig:SM}
\end{figure}

As illustrated in Fig.~\ref{fig:SM}, we consider a non-terrestrial satellite communication system consisting of a LEO satellite, a UAV-mounted BD-ARIS with $M$ elements, and $I$ ground users, denoted by $\{U_1, U_2, \dots, U_I\}$, located at distinct horizontal positions on the Earth's surface.

The regenerative satellite, orbiting at an altitude of 520~km, is equipped with a uniform rectangular array (URA) comprising $N$ active transmit antennas. It serves as the primary signal source and provides downlink connectivity using RSMA, a robust multiple access technique that splits each user's message into a common part and a private part, enabling flexible interference management and simultaneous transmission to multiple users over the same frequency-time resources.

To enhance signal quality and extend coverage, a BD-ARIS is deployed on a UAV. The UAV maintains a fixed altitude $h_{\text{UAV}}$ while \textcolor{black}{repositioning its horizontal location in the ($xy$)-plane according to the current network conditions.} The BD-ARIS consists of $M$ active elements capable of imposing adjustable phase shifts and amplifying the incident signals. These elements are structured in a coupled configuration to capture inter-element interactions. The UAV carries the BD-ARIS, which receives the incident electromagnetic wave from the satellite and re-radiates its amplified and spatially
transformed signal toward the ground users.

The UAV's position is represented as $\mathbf{Pos}_{\text{UAV}} = [x_{\text{UAV}}, y_{\text{UAV}}, h_{\text{UAV}}]^T$, where $x_{\text{UAV}}$ and $y_{\text{UAV}}$ denote the horizontal coordinates, and $h_{\text{UAV}}$ is the fixed altitude. To maximize the system's EE, the UAV dynamically determines its optimal horizontal location $(x_{\text{UAV}}^*, y_{\text{UAV}}^*)$ based on slow-varying network conditions, such as users' rate requirements and channel states. \textcolor{black}{After reaching the optimized horizontal location, the UAV remains stationary during the corresponding communication/CSI-update interval. Hence, the present formulation optimizes UAV placement rather than continuous UAV trajectory, and horizontal motion during active transmission is not modeled.}

\subsection{Channels Model}
\label{sec:channel_model}

We consider a downlink RSMA system consisting of a LEO satellite, a UAV-mounted BD-ARIS, and $I$ single-antenna ground users. The baseband channel variables are denoted as
\(
\mathbf{h}_i \in \mathbb{C}^{N\times1},
\quad
\mathbf{g}_i(\mathbf{Pos}_{\mathrm{UAV}}) \in \mathbb{C}^{M\times1},
\quad
\mathbf{H}_u(\mathbf{Pos}_{\mathrm{UAV}}) \in \mathbb{C}^{M\times N}
\).
The dependence of $\mathbf{g}_i$ and $\mathbf{H}_u$ on $\mathbf{Pos}_{\mathrm{UAV}}$ reflects the fact that UAV repositioning affects path loss and elevation/azimuth angles. These, in turn, determine the spatial signatures observed at the BD-ARIS.

\textcolor{black}{For satellite-related links, the physical and effective channel representations are distinguished below.}

\subsubsection*{1) Rician fading model}

All wireless links follow a Rician fading model:
\begin{equation}
\mathbf{X}(\mathbf{Pos}_{\mathrm{UAV}})
=
\sqrt{\frac{K_{\mathbf{X}}}{K_{\mathbf{X}}+1}}\,
\mathbf{X}^{\mathrm{LoS}}
+
\sqrt{\frac{1}{K_{\mathbf{X}}+1}}\,
\mathbf{X}^{\mathrm{NLoS}},
\label{eq:rician_fading}
\end{equation}
where $K_{\mathbf{X}}$ is the Rician $K$-factor. The NLoS component is modeled as
\(
\mathbf{X}^{\mathrm{NLoS}}(\mathbf{Pos}_{\mathrm{UAV}})
\sim
\mathcal{CN}\!\left(
\mathbf{0},
\beta_0 d_{\mathbf{X}}^{-\ell}\mathbf{I}
\right),
\)
where $d_{\mathbf{X}}$ is the geometric distance of the corresponding link and $\ell$ is the path-loss exponent. For satellite-related links, the distance and the corresponding NLoS channel component are time-varying through the LEO trajectory; for static links, the time argument is omitted for notational simplicity.

\subsubsection*{2) LoS components and geometric dependence}

\paragraph{Satellite--User LoS}
\begin{equation}
{\color{black}
\begin{aligned}
\mathbf{h}_{i}^{\mathrm{LoS}}(t)
&=
\sqrt{G_{\mathrm{sat}}G_i}
\left(\frac{c}{4\pi f_c d_{s,i}(t)}\right)^{\ell/2}\times
\mathbf{a}_{\mathrm{SAT}}\!\left(\theta_i(t),\phi_i(t)\right)
,
\end{aligned}
}
\label{eq:los_sat_user_final}
\end{equation}
\textcolor{black}{where $\mathbf{h}_{i}^{\mathrm{LoS}}(t)$ denotes the time-varying LoS component, $d_{s,i}(t)$ is the instantaneous satellite--user distance, $G_{\mathrm{sat}}$ and $G_i$ are the antenna gains, and $\mathbf{a}_{\mathrm{SAT}}(\theta_i(t),\phi_i(t))$ is the time-varying satellite array steering vector. The time-varying propagation-phase component is represented through the Doppler-induced phase in \eqref{eq:phy_sat_user}.}

\textcolor{black}{Under the adopted narrowband far-field assumption, the Doppler effect is modeled as a common phase rotation across the satellite antenna array, while the spatial phase differences across the array are captured by the corresponding time-varying steering vector.}

\paragraph{Satellite--UAV LoS (Satellite$\rightarrow$RIS)}

The LoS channel from the satellite to the RIS mounted on the UAV is
\begin{equation}
{\color{black}
\begin{aligned}
\mathbf{H}_{u}^{\mathrm{LoS}}(t)
&=
\sqrt{\beta_0\,d_{s,u}(t)^{-\ell}}\,
\mathbf{a}_{\mathrm{RIS}}\!\left(\theta_u(t),\phi_u(t)\right)
\\[-1mm]
&\quad\times
\mathbf{a}_{\mathrm{SAT}}^{H}\!\left(\theta_s(t),\phi_s(t)\right),
\end{aligned}
}
\label{eq:los_sat_uav_final}
\end{equation}
\textcolor{black}{where $\mathbf{H}_{u}^{\mathrm{LoS}}(t)$ denotes the time-varying LoS component of the satellite--UAV channel and $d_{s,u}(t)$ is the instantaneous satellite--UAV distance. The angles $(\theta_u(t),\phi_u(t))$ and $(\theta_s(t),\phi_s(t))$ characterize the corresponding elevation and azimuth directions at the UAV-RIS and satellite, respectively.}

\paragraph{UAV--User LoS (RIS$\rightarrow$User)}
\begin{equation}
\mathbf{g}_i^{\mathrm{LoS}}(\mathbf{Pos}_{\mathrm{UAV}})
=
\sqrt{\beta_0\, d_{u,i}^{-\ell}}\;
\mathbf{a}_{\mathrm{RIS}}(\theta_{u\to i},\phi_{u\to i}),
\label{eq:los_uav_user_final}
\end{equation}
where ($d_{u,i}$) is the UAV--user distance, with ($\theta_{u\to i}$) being the elevation angle between the UAV-user line and the horizontal plane, and ($\phi_{u\to i}$) representing the azimuth angle from the positive x-axis to the projection of the UAV-user vector.

\textcolor{black}{The complete physical channels of the satellite--user and satellite--UAV links are given by}
\begin{equation}
{\color{black}
\begin{aligned}
\mathbf{h}_{i,\mathrm{phy}}(t)
&=
\left[
\sqrt{\frac{K_{s,i}}{K_{s,i}+1}}\,
\mathbf{h}_{i}^{\mathrm{LoS}}(t)
\right.
\\[-1mm]
&\quad\left.
+\sqrt{\frac{1}{K_{s,i}+1}}\,
\mathbf{h}_{i}^{\mathrm{NLoS}}(t)
\right]
e^{j\phi_{D,s,i}(t)},
\end{aligned}
}
\label{eq:phy_sat_user}
\end{equation}
and
\begin{equation}
{\color{black}
\begin{aligned}
\mathbf{H}_{u,\mathrm{phy}}(t)
&=
\left[
\sqrt{\frac{K_{s,u}}{K_{s,u}+1}}\,
\mathbf{H}_{u}^{\mathrm{LoS}}(t)
\right.
\\[-1mm]
&\quad\left.
+\sqrt{\frac{1}{K_{s,u}+1}}\,
\mathbf{H}_{u}^{\mathrm{NLoS}}(t)
\right]
e^{j\phi_{D,s,u}(t)}.
\end{aligned}
}
\label{eq:phy_sat_uav}
\end{equation}

\textcolor{black}{The Doppler-induced phases $\phi_{D,s,i}(t)$ and $\phi_{D,s,u}(t)$ are defined in Subsection~II.A-4}.

\textcolor{black}{To distinguish the physical channels from the channels used for resource optimization, the compensated effective channels are defined as}
\begin{equation}
{\color{black}
\bar{\mathbf{h}}_i(t)
=
\mathbf{h}_{i,\mathrm{phy}}(t)
e^{-j\widehat{\phi}_{D,s,i}(t)},
}
\label{eq:effective_sat_user}
\end{equation}
and
\begin{equation}
{\color{black}
\bar{\mathbf{H}}_u(t)
=
\mathbf{H}_{u,\mathrm{phy}}(t)
e^{-j\widehat{\phi}_{D,s,u}(t)},
}
\label{eq:effective_sat_uav}
\end{equation}
\textcolor{black}{where $\widehat{\phi}_{D,s,i}(t)$ and $\widehat{\phi}_{D,s,u}(t)$ are the estimated Doppler-induced phases. If the Doppler estimate is imperfect, a residual phase remains in the effective channel. For notational simplicity, the overbar is omitted hereafter, and $\mathbf{h}_i$ and $\mathbf{H}_u$ denote the effective post-compensation channels used in the signal model and optimization framework.}

\textcolor{black}{In the considered system, the effective CSI is acquired through a periodic pilot-assisted estimation procedure after compensation of the predictable nominal Doppler component. At the beginning of each CSI update interval of duration $T_{\mathrm{CSI}}$, the satellite state and relative propagation geometry are used to estimate the nominal Doppler shift, and the corresponding phase/frequency offset is compensated through transmitter-side pre-compensation and/or receiver-side tracking. Pilot signals are then used to estimate the effective satellite--user, satellite--UAV, and UAV--user channels required for RSMA precoding and BD-ARIS configuration.}

\textcolor{black}{The CSI update interval $T_{\mathrm{CSI}}$ is selected according to the temporal variation of the residual effective channel after nominal Doppler compensation, such that}
\(
T_{\mathrm{CSI}} \leq T_{\mathrm{c,eff}},
\)
\textcolor{black}{where $T_{\mathrm{c,eff}}$ denotes the effective coherence interval of the post-compensation channel. Thus, the communication controller refreshes the estimated effective CSI at the beginning of each update interval and keeps the corresponding control action fixed during that interval unless a new CSI update is triggered.}

\subsubsection*{3) CSI uncertainty}

To model practical estimation errors, we adopt an additive Gaussian uncertainty model,
\(
\mathbf{X}
=
\widehat{\mathbf{X}} + \Delta\mathbf{X},
\quad
\Delta\mathbf{X}\sim
\mathcal{CN}\!\left(\mathbf{0},\sigma_{\mathbf{X}}^{2}\mathbf{I}\right),
\)
where ($\widehat{\mathbf{X}}$) is the estimated channel and ($\sigma_{\mathbf{X}}^{2}$) decreases with the pilot SNR.

\textcolor{black}{In practical LEO links, the estimated effective CSI may still contain residual errors caused by imperfect Doppler estimation and tracking, synchronization and timing errors, pilot contamination, limited feedback, and quantization. Accordingly, $\Delta\mathbf{X}$ is interpreted as an aggregate residual CSI error that captures these practical impairments, including the residual Doppler mismatch $f_{D,q}(t)-\widehat{f}_{D,q}(t)$. The adopted Gaussian model is therefore an aggregate system-level approximation: individual impairment mechanisms are not separately parameterized, and their combined residual effect is represented through the effective error variance $\sigma_{\mathbf{X}}^{2}$.}

\textcolor{black}{Thus, the optimization and DRL framework operates on estimated effective CSI rather than ideal physical CSI.}

The model in \eqref{eq:rician_fading}--\eqref{eq:los_uav_user_final}, together with \eqref{eq:phy_sat_user}--\eqref{eq:effective_sat_uav}, captures the UAV-position-dependent channel characteristics, the BD-ARIS spatial transformation, and practical CSI uncertainty.

\subsubsection*{4) Impact of LEO Satellite Mobility on Channels}
\label{subsec:mobility}

The high orbital velocity of the LEO satellite allows it to sweep a wide ground area, thereby providing coverage to users located within its field of view. In this work, we consider a finite time window during which the satellite is visible to both the UAV and the ground users, and we model its trajectory as a circular orbit.

{\color{black}
Due to the orbital motion, the satellite--UAV and satellite--user links are inherently time-varying. For tractability, the satellite trajectory over the considered visibility window is modeled as a three-dimensional circular orbit in a common Cartesian reference frame. Let the satellite position be
\begin{equation}
\mathbf{p}_{\mathrm{sat}}(t)
=
R_{\mathrm{orb}}
\begin{bmatrix}
\cos\big(\nu(t)\big)\\
\sin\big(\nu(t)\big)\cos(i_{\mathrm{orb}})\\
\sin\big(\nu(t)\big)\sin(i_{\mathrm{orb}})
\end{bmatrix},
\label{eq:satellite_trajectory}
\end{equation}
where
\(
R_{\mathrm{orb}}=R_{\mathrm{E}}+h_{\mathrm{sat}},
\quad
\nu(t)=\nu_0+\omega_{\mathrm{orb}}t, \) with $R_{\mathrm{E}}$ denoting the Earth radius, $h_{\mathrm{sat}}$ the LEO orbital altitude, $i_{\mathrm{orb}}$ the orbital inclination, $\nu_0$ the initial true anomaly, and $\omega_{\mathrm{orb}}$ the angular orbital velocity. The UAV and ground-user positions are represented in the same Cartesian reference frame when evaluating the instantaneous propagation distances.} The corresponding distances are
\begin{align}
d_{s,u}(t) &= \big\|
\mathbf{p}_{\mathrm{sat}}(t) - \mathbf{p}_{\mathrm{UAV}}
\big\|,\\
d_{s,i}(t) &= \big\|
\mathbf{p}_{\mathrm{sat}}(t) - \mathbf{p}_{\mathrm{U}_i}
\big\|.
\end{align}

The elevation and azimuth angles in the satellite-related LoS channels are consequently time-varying functions of the instantaneous satellite and receiver positions.

\textcolor{black}{The resulting instantaneous Doppler shift is determined by the temporal variation of the propagation distance as}
\begin{equation}
{\color{black}
f_{D,q}(t)
=
-\frac{1}{\lambda_c}
\frac{d d_q(t)}{dt},
\qquad
\lambda_c=\frac{c}{f_c},
\qquad
q\in\{(s,u),(s,i)\},
}
\label{eq:general_doppler}
\end{equation}
\textcolor{black}{where $d_q(t)$ denotes the corresponding satellite--UAV or satellite--user distance. The accumulated Doppler-induced phase used in \eqref{eq:phy_sat_user} and \eqref{eq:phy_sat_uav} is therefore}
\begin{equation}
{\color{black}
\phi_{D,q}(t)
=
2\pi\int_{0}^{t} f_{D,q}(\tau)\,d\tau.
}
\label{eq:doppler_phase}
\end{equation}

\textcolor{black}{Since the Doppler shift is caused by the relative motion of the satellite and the corresponding communicating node, it is directly determined by the time-varying satellite trajectory and propagation distance. The nominal Doppler component can therefore be estimated from the satellite state and relative propagation geometry and compensated by the communication processor through Doppler pre-compensation and/or receiver-side tracking.}

Consequently, both the LoS components and the NLoS components become time-dependent, resulting in time-varying satellite--RIS--user cascaded channels.

\textcolor{black}{The residual mismatch after Doppler compensation is incorporated into the imperfect-CSI model described above.}

\subsection{BD-Active RIS with Group-Connected Architecture}

In an {active} BD-RIS, each element not only applies a controllable phase/amplitude adjustment but can also {amplify} the impinging signal through low-noise amplification circuitry. Moreover, due to the BD structure, neighboring elements within each group exhibit \emph{mutual coupling}, which results in a more general and practical model compared with conventional diagonal (element-wise independent) RISs.

In the group-connected architecture with group size 2, the RIS contains
\( G=\frac{M}{2} \) groups (assuming even \(M\)), where each group consists of two coupled and actively-amplified elements. The overall scattering matrix of the BD-ARIS is modeled as a block-diagonal matrix:
\begin{equation}
\boldsymbol{\Phi} = \mathrm{diag}(\boldsymbol{\Phi}_{1}, \boldsymbol{\Phi}_{2}, \dots, \boldsymbol{\Phi}_{G}),
\end{equation}
where each group matrix \( \boldsymbol{\Phi}_{g} \in \mathbb{C}^{2 \times 2} \) has the symmetric form:
\begin{equation}
\boldsymbol{\Phi}_{g} =
\begin{bmatrix}
\phi_{g,1} & b_g \\
b_g & \phi_{g,2}
\end{bmatrix},
\label{eq:phi_block}
\end{equation}
with \( \boldsymbol{\Phi}_{g} = \boldsymbol{\Phi}_{g}^T \) due to the implemented symmetric coupling network.
Here, \( \phi_{g,1}=\beta_{g,1}e^{j\theta_{g,1}} \) and \( \phi_{g,2}=\beta_{g,2}e^{j\theta_{g,2}} \) denote the complex (amplitude--phase) reflection--amplification coefficients of the two active elements in group \(g\), while \( b_g \) represents the complex-valued inter-element coupling coefficient. In practical hardware, \(b_g\) can be real-valued or conjugate-symmetric, depending on the coupling implementation.

Unlike passive BD-RIS architectures, where the constraint
$\boldsymbol{\Phi}_g \boldsymbol{\Phi}_g^H \preceq \mathbf{I}$ enforces energy preservation,
the matrices in \eqref{eq:phi_block} here represent \emph{active} amplification circuitry.
Therefore, the spectral norm of $\boldsymbol{\Phi}_{g}$ is not restricted to unity,
and amplification gains $\beta_{g,i} > 1$ are permitted.
As in all active architectures, signal amplification intrinsically introduces
\emph{internal noise}, which will be modeled explicitly in the received signal
and in the RIS power constraints in the next subsection.

This group-based active modeling captures (i) mutual coupling, (ii) signal amplification, and (iii) active hardware behavior with significantly lower implementation overhead compared with fully non-diagonal architectures. The BD-ARIS is mounted on the UAV, and because its configuration depends on the UAV's horizontal position, the system adaptively reinforces the downlink as \textcolor{black}{network conditions evolve}, enabling improved coverage and rate performance. \textcolor{black}{\footnote{From a practical deployment perspective, the UAV-mounted BD-ARIS requires that the panel weight and power be compatible with the UAV's payload. Recent lightweight metasurface designs and low-power active components make RIS panels with several tens of elements feasible on mid-size UAVs. The RIS is remotely configured via a low-rate control link; the sub-millisecond inference time of DG-TRPO allows the UAV to update the RIS matrix well within the LEO channel coherence time.}}

\subsection{Transmission Protocol}

In the considered system, RSMA is adopted to simultaneously serve $I$ ground users. The satellite transmits a signal comprising a common message $s_c$, intended for all users, and $I$ private messages $\{s_i\}_{i=1}^{I}$, each intended for a particular user. The transmitted signal is modeled as \cite{10915662}:
\begin{equation}
\mathbf{x}_s =
\sqrt{P_s a_c}\, \mathbf{w}_c s_c
+
\sum_{i=1}^{I}
\sqrt{P_s a_i}\, \mathbf{w}_i s_i,
\label{eq:transmitted_signal}
\end{equation}
where $P_s$ is the total satellite transmit power, $a_c,a_i\in[0,1]$ are power allocation coefficients satisfying
$a_c + \sum_{i=1}^{I} a_i = 1$,
$\mathbf{w}_c,\mathbf{w}_i \in \mathbb{C}^{N\times1}$,
$\|\mathbf{w}_c\|^2 = \|\mathbf{w}_i\|^2 =1$
are the RSMA precoders, and $s_c, s_i$ are unit-power data symbols.

The received signal at user $U_i$ consists of the direct satellite link and the UAV-mounted BD-ARIS link. Because each group of the BD-ARIS contains active amplification circuitry, it injects \emph{amplified internal noise} into the cascaded link. Consequently, the received baseband signal is
\begin{align}
y_i &= \mathbf{h}_i^H \mathbf{x}_s
      + \mathbf{g}_i^H(\mathrm{Pos}_{\mathrm{UAV}})\, \boldsymbol{\Phi}\,
        \mathbf{H}_u(\mathrm{Pos}_{\mathrm{UAV}})\, \mathbf{x}_s \notag \\
    &\quad + \mathbf{g}_i^H(\mathrm{Pos}_{\mathrm{UAV}})\, \mathbf{n}_{\mathrm{RIS}} + n_i,
\label{eq:received_signal_active}
\end{align}
where $n_i \sim \mathcal{CN}(0,\sigma_i^2)$ is the user noise and
$\mathbf{n}_{\mathrm{RIS}} \sim \mathcal{CN}(\mathbf{0},\Sigma_{\mathrm{RIS}})$
is the aggregated amplified noise generated by all active RIS groups,
with block-diagonal covariance
$\Sigma_{\mathrm{RIS}}=\mathrm{diag}(\Sigma_{\mathrm{RIS},1},\ldots,\Sigma_{\mathrm{RIS},G})$.
The noise covariance of the $g$-th group is modeled as
$\Sigma_{\mathrm{RIS},g} = \sigma_{\mathrm{amp}}^2 \big(|\phi_{g,1}|^2 + |\phi_{g,2}|^2 + 2|b_g|^2\big)\,\mathbf{I}_2$,
where $\sigma_{\mathrm{amp}}^2$ denotes the amplifier noise variance,
capturing the dependence on the element gains and inter-element coupling.
Using the structure of the BD-ARIS-assisted link, the equivalent channel seen by user $U_i$ is defined by:
\begin{equation}
\mathbf{H}_{\mathrm{eq},i}(\mathrm{Pos}_{\mathrm{UAV}})
=
\mathbf{h}_i^H
+
\mathbf{g}_i^H(\mathrm{Pos}_{\mathrm{UAV}})\,
\boldsymbol{\Phi}\,
\mathbf{H}_u(\mathrm{Pos}_{\mathrm{UAV}}),
\label{eq:equivalent_channel_active}
\end{equation}
which captures both the direct path and the actively amplified UAV-assisted path.
The UAV position
$\mathrm{Pos}_{\mathrm{UAV}} = (x_{\mathrm{UAV}},y_{\mathrm{UAV}},h_{\mathrm{UAV}})$
affects both $\mathbf{g}_i$ and $\mathbf{H}_u$, enabling adaptive reinforcement of user links.
Each user first decodes the common message $s_c$ by treating private streams as interference. Due to the presence of amplified RIS noise, the SINR for decoding $s_c$ at user $U_i$ becomes
\begin{equation}
\gamma_{c,i}
=
\frac{P_s a_c \left| \mathbf{H}_{\mathrm{eq},i} \mathbf{w}_c \right|^2}
{\sum_{j=1}^{I} P_s a_j \left| \mathbf{H}_{\mathrm{eq},i} \mathbf{w}_j \right|^2
\;+\;
\mathbf{g}_i^H \Sigma_{\mathrm{RIS}} \mathbf{g}_i
\;+\;
\sigma_i^2 }.
\label{eq:sinr_common_active}
\end{equation}

After successfully decoding and canceling the common message via SIC, user $U_i$ decodes its private message with the private-stream SINR:
\begin{equation}
\gamma_{p,i}
=
\frac{P_s a_i \left| \mathbf{H}_{\mathrm{eq},i} \mathbf{w}_i \right|^2}
{\sum_{j\ne i} P_s a_j \left| \mathbf{H}_{\mathrm{eq},i} \mathbf{w}_j \right|^2
\;+\;
\mathbf{g}_i^H \Sigma_{\mathrm{RIS}} \mathbf{g}_i
\;+\;
\sigma_i^2 }.
\label{eq:sinr_private_active}
\end{equation}

The achievable common rate at user $U_i$ is
\begin{equation}
R_{c,i} = B\log_2(1 + \gamma_{c,i}),
\end{equation}
and the actual common rate is
\begin{equation}
R_c = \min_{i\in\{1,\ldots,I\}} R_{c,i}.
\end{equation}
If user $U_i$ is assigned a fraction $\delta_i$ of the common rate, its
individual achievable rate is
\begin{equation}
R_i
=
\delta_i R_c
+
B\log_2(1+\gamma_{p,i}),
\end{equation}
where $\sum_{i=1}^{I}\delta_i=1$. Therefore, the total achievable sum rate is
\begin{equation}
R_{\Sigma}
=
\sum_{i=1}^{I}R_i
=
R_c+B\sum_{i=1}^{I}\log_2(1+\gamma_{p,i}).
\label{eq:sum_rate}
\end{equation}

This transmission protocol fully incorporates the effects of the UAV-mounted BD-ARIS: amplification, coupling, and internal amplified noise. The resulting equivalent channel and SINR expressions capture the joint impact of direct satellite links and actively enhanced UAV-assisted links, yielding a flexible and robust multi-user downlink transmission scheme.

\subsection{Power Consumption Model}

In this section, we analyze the total power consumption of the UAV-mounted BD-ARIS system,
which includes several key components: the satellite's transmission power, the amplification power
of RIS elements, the signal processing power at the UAV, and the UAV's mechanical power for hovering.
The total power consumption is expressed as:
\begin{equation}
P_{\text{total}} =  P_{\text{s}} + P_{\text{BD-ARIS}} + P_{\text{proc}} + P_{\text{UAV}},
\label{eq:total_power}
\end{equation}
where \(P_{\text{s}}\) is the total satellite
transmit power.\textcolor{black}{\footnote{Because the RSMA power allocation coefficients satisfy
$ a_c + \sum_{i=1}^I a_i = 1 $ by constraint~\eqref{eq:c5_active}, the total radiated power
equals $P_s$, and the individual contributions of the common and private streams are governed by
$a_c$ and $\{a_i\}$.}} 
Furthermore, \( P_{\text{BD-ARIS}} \) accounts for the power consumed by the BD-ARIS
elements for signal amplification and reflection, \( P_{\text{proc}} \) denotes the
processing power required for baseband operations and control signaling at the UAV, and
\( P_{\text{UAV}} \) represents the mechanical power required to maintain the UAV's stable
hovering at the designated altitude.

\subsubsection{Power Consumption of BD-Active RIS}
To calculate the total power consumption in the BD-ARIS system with a group size of 2, we consider the power consumed by the signal amplification, phase shifters, and DC biasing. The total power consumption can be expressed as:
\begin{equation}
    P_{\mathrm{BD-ARIS}} = \vartheta_{\mathrm{RIS}} P_{\mathrm{out}} + \frac{M}{2} \left( P^{\mathrm{RIS}}_{\mathrm{D}} + P^{\mathrm{RIS}}_{\mathrm{DC}} \right),
\end{equation}
where \( \vartheta_{\mathrm{RIS}} \) denotes the inverse power-amplifier
efficiency and \( P_{\mathrm{out}} \) is the total output power transmitted
by the BD-ARIS to all users. The power consumption associated with each
group/amplification branch includes both phase-shifter and DC-biasing
components: \( P^{\mathrm{RIS}}_{\mathrm{D}} \) depends on the resolution
(e.g., 1.5--7.8 mW for 3--6 bits \cite{huang2019reconfigurable}), and
\( P^{\mathrm{RIS}}_{\mathrm{DC}} \) accounts for element control
\cite{niu2023active}. With a group size of 2, each amplifier serves two
elements, reducing the total power consumption compared to conventional
active RIS designs.

Since the number of two-element groups is $G=M/2$, the maximum allowable
output power of each group is scaled with the number of RIS elements as
\begin{equation}
P_{\mathrm{RIS},g}^{\max}
=
\frac{\kappa}{G}P_{\mathrm{RIS}}^{\max}
=
\frac{2\kappa}{M}P_{\mathrm{RIS}}^{\max},
\end{equation}
where $P_{\mathrm{RIS}}^{\max}$ denotes the total RIS output-power budget
and $\kappa$ is a group-power scaling factor. 

\subsubsection{Hovering Power Consumption Model for Rotary-Wing UAVs}

In hovering mode, the propulsion power consumption of a rotary-wing UAV mainly consists of two components: the blade profile power (\( P_0 \)) and the induced power required to produce lift in hover (\( P_i \)) \cite{zeng2019energy}. The total required power is:
\begin{equation}
    P_{\mathrm{h}} = 
    \underbrace{\frac{\delta}{8} \rho s A \Omega^3 R^3}_{P_0} 
    + 
    \underbrace{(1 + k) \frac{W^{3/2}}{\sqrt{2 \rho A}}}_{P_i},
\end{equation}
here, \( W \) denotes the UAV weight, \( \rho \) is the air density (kg/m\textsuperscript{3}), \( s \) is the rotor solidity, \( A = \pi R^2 \) is the rotor disc area, \( R \) is the rotor radius (m), \( \Omega \) is the angular velocity (rad/s), \( \delta \) is the blade profile drag coefficient, and \( k \) is the induced power correction factor. This expression provides an analytical model for the hovering power, which plays a fundamental role in evaluating the EE of UAV-enabled communication systems. 

Additionally, the UAV consumes processing power \(P_{\text{proc}}\) (1--5 W typ.) for channel estimation and beam control. \textcolor{black}{In the present formulation, the UAV is assumed to remain stationary during each communication/CSI-update interval; therefore, the propulsion energy associated with horizontal repositioning or continuous trajectory motion is not modeled. The UAV mechanical power included in the EE objective is thus the hovering power required to maintain the selected operating position.}

\section{Problem Formulation}
\label{sec:problem_formulation}

The main objective of this paper is to maximize the overall EE of a LEO satellite
communication system assisted by a UAV-mounted group-connected BD-ARIS.
The optimization problem is formulated to maximize the sum of the achievable common
and private rates for all \( I \) users under the RSMA scheme, normalized by the
total power consumption. Specifically, the UAV's position \( \mathbf{Pos}_{\mathrm{UAV}} \)
is optimized to maximize:

\begin{equation}
\mathrm{EE} =
\frac{R_{\Sigma}\big(a_c, \{a_i\}, \mathbf{w}_c, \{\mathbf{w}_i\},
\boldsymbol{\Phi}, \mathbf{Pos}_{\mathrm{UAV}}\big)}
{P_{\text{total}}\big(a_c, \{a_i\}, \mathbf{w}_c, \{\mathbf{w}_i\},
\boldsymbol{\Phi}, \mathbf{Pos}_{\mathrm{UAV}}\big)},
\label{eq:EE_def}
\end{equation}
where the total power consumption \( P_{\text{total}} \) is given
by~\eqref{eq:total_power}. The sum rate $R_\Sigma$ is computed by including
the system bandwidth $B$, so the EE defined in~\eqref{eq:EE_def} is expressed
in bit/J. In the simulations, all reported EE values are given in Kbits/Joule.

\begin{subequations}
\label{eq:opt_problem_RSMA}
\begin{align}
\max_{\substack{
a_c,\{a_i\},\mathbf{w}_c,\{\mathbf{w}_i\},\\
\boldsymbol{\Phi},x_{\mathrm{UAV}},y_{\mathrm{UAV}}
}}
\quad &
\mathrm{EE}(a_c,\{a_i\},\mathbf{w}_c,\{\mathbf{w}_i\},\boldsymbol{\Phi},x_{\mathrm{UAV}},y_{\mathrm{UAV}})
\label{eq:opt_problem_RSMA_obj_active}
\\[1mm]
\text{s.t.}\quad &
\gamma_c = \min_{i} \gamma_{c,i} \ge \gamma_{\min}^{(c)}, 
\label{eq:c1_active}
\\
& \gamma_{p,i} \ge \gamma_{\min}^{(i)},\quad \forall i,
\label{eq:c2_active}
\\
& 0 \le P_s(a_c + \sum_{i} a_i) \le P_{\mathrm{SAT}}^{\max},
\label{eq:c3_active}
\\
& P_{\mathrm{out},g}(\boldsymbol{\Phi}_g)
\le P_{\mathrm{RIS},g}^{\max},\quad \forall g,
\label{eq:c4_active_group}
\\
& P_{\mathrm{out}}
\le P_{\mathrm{RIS}}^{\max},
\label{eq:c4_active_total}
\\
& a_c + \sum_{i=1}^I a_i = 1,
\label{eq:c5_active}
\\
& 0 \le a_c, a_i \le 1, \quad \forall i,
\label{eq:c6_active}
\\
& \|\mathbf{w}_c\|^2 =  \|\mathbf{w}_i\|^2  = 1, \label{eq:c7_active_w_norm} \\
& \boldsymbol{\Phi}_g = \boldsymbol{\Phi}_g^T, \quad \forall g,
\label{eq:c7_active_symmetry}
\\
& x_{\mathrm{UAV}} \le x_{\max}, \quad
  y_{\mathrm{UAV}} \le y_{\max},
\label{eq:c9_active}
\\
& x_{\mathrm{UAV}}, y_{\mathrm{UAV}} \ge 0.
\label{eq:c10_active}
\end{align}
\end{subequations}

In \eqref{eq:opt_problem_RSMA}, constraint~\eqref{eq:c1_active} ensures
reliable decoding of the common stream by all users through the minimum
common-stream SINR requirement, while constraint~\eqref{eq:c2_active}
guarantees the individual private-stream QoS requirements after SIC.
Constraint~\eqref{eq:c3_active} limits the total satellite transmit power
to the available satellite power budget.

Constraints~\eqref{eq:c4_active_group} and~\eqref{eq:c4_active_total}
impose the per-group and total output-power limits of the BD-ARIS,
respectively. For each group $g$, the corresponding output power is
\(
P_{\mathrm{out},g}(\boldsymbol{\Phi}_g)
=
\mathrm{tr}\!\left(
\boldsymbol{\Phi}_g\,\mathbf{C}_{\mathrm{in},g}\,
\boldsymbol{\Phi}_g^{H}
\right)
+
\mathrm{tr}(\Sigma_{\mathrm{RIS},g}),
\)
where $\mathbf{C}_{\mathrm{in},g}$ denotes the incident-signal covariance
matrix corresponding to the $g$-th group. The total BD-ARIS output power
is defined as
\(
P_{\mathrm{out}}
=
\sum_{g=1}^{G}P_{\mathrm{out},g}.
\)
The per-group limit $P_{\mathrm{RIS},g}^{\max}$ is determined according
to the group-power scaling rule introduced in the power-consumption model,
while $P_{\mathrm{RIS}}^{\max}$ denotes the total BD-ARIS output-power
budget. If the per-group limits were set to sum exactly to
$P_{\mathrm{RIS}}^{\max}$, constraint~\eqref{eq:c4_active_group} would
become redundant with respect to~\eqref{eq:c4_active_total}. To avoid this
degeneracy and keep both constraints simultaneously active, the
group-power scaling rule adopts $\kappa=2$, so that each group may draw up
to twice its equal-share allocation while the aggregate RIS output power
remains bounded by $P_{\mathrm{RIS}}^{\max}$.

Constraint~\eqref{eq:c5_active}, together with~\eqref{eq:c6_active},
defines the RSMA power-allocation coefficients, whereas
\eqref{eq:c7_active_w_norm} imposes unit-norm constraints on the
beamforming vectors so that the allocated coefficients directly control
the satellite transmit power in~\eqref{eq:transmitted_signal}.
Constraint~\eqref{eq:c7_active_symmetry} enforces the symmetric structure
of each two-element BD-ARIS group, while
constraints~\eqref{eq:c9_active}--\eqref{eq:c10_active} define the feasible
horizontal placement region of the UAV.

Due to the non-convexity and tightly coupled constraints
in~\eqref{eq:opt_problem_RSMA}, real-time solution is challenging. We first
develop a model-based iterative algorithm as an offline model-based
benchmark, and then propose two learning-based schemes, standard TRPO and
the novel Dinkelbach-Guided TRPO, that achieve high EE with
sub-millisecond inference, robust to CSI errors and satellite mobility.

\section{Model-Based Optimization via Dinkelbach, WMMSE, SCA, and SQP}
\label{sec:classical_solution}

The EE maximization problem~\eqref{eq:opt_problem_RSMA} is a non-convex fractional
program with strongly coupled variables. We propose an iterative algorithm combining
Dinkelbach's transform, alternating optimization (AO), WMMSE rate reformulation,
successive convex approximation (SCA) for the BD-ARIS blocks, and sequential
quadratic programming (SQP) for UAV placement.

\subsection{Fractional Programming via Dinkelbach}
Let $\mathbf{x} \triangleq \big( a_c, \{a_i\}, \mathbf{w}_c, \{\mathbf{w}_i\},
\boldsymbol{\Phi}, \mathbf{Pos}_{\mathrm{UAV}} \big)$, define
$f(\mathbf{x}) \triangleq R_{\Sigma}(\mathbf{x})$ and
$g(\mathbf{x}) \triangleq P_{\mathrm{total}}(\mathbf{x})$.
The original fractional
objective is $\mathrm{EE}(\mathbf{x}) = f(\mathbf{x})/g(\mathbf{x})$. Dinkelbach's
method transforms EE maximization into a sequence of parameterized subproblems
\begin{equation}
\mathbf{x}^{\star}(\mu)=\arg\max_{\mathbf{x}\in\mathcal{X}} \;
F(\mathbf{x};\mu)\coloneqq f(\mathbf{x})-\mu\,g(\mathbf{x}),
\label{eq:dinkelbach_subproblem}
\end{equation}
with iterative update $\mu\leftarrow f(\mathbf{x}^\star)/g(\mathbf{x}^\star)$ until
$|f-\mu g|\le\epsilon_{\mathrm{D}}$. A few ($5\!-\!15$) outer iterations suffice. The
feasible set $\mathcal{X}$ collects constraints~\eqref{eq:c1_active}--\eqref{eq:c10_active}.

\subsection{WMMSE Reformulation of the Achievable Rates}

To handle the non-convex rate expressions, we apply the WMMSE relationship. For any
stream with SINR $\gamma_i$,
\(
\ln(1+\gamma_i)
=
\max_{\omega_i,u_i}
\left(
\ln\omega_i-\omega_i e_i(u_i)+1
\right),
\)
where $u_i$ is the linear receiver,
$e_i(u_i)=|u_i^*y_i-s_i|^2$ is the MSE, and $\omega_i$ is the weight.
This transformation is applied to both the common and private streams.

For the common stream $s_c$, an auxiliary natural-log common-rate variable
$\bar{r}_c$ is introduced with
$\bar{r}_c \le \ln(1+\gamma_{c,i})$ for all $i$. The corresponding sum rate is represented as
$R_{\Sigma} = \frac{B}{\ln 2}\left(\bar{r}_c +
\sum_i \ln(1+\gamma_{p,i})\right)$. The inner Dinkelbach subproblem then reads
\begin{equation}
\max_{\substack{a_c,\{a_i\},\mathbf{w},\boldsymbol{\Phi},\mathbf{u},\omega,\\
\bar{r}_c,\mathbf{x}_{\mathrm{UAV}}}} \;
\frac{B}{\ln 2}\Big(\bar{r}_c + \sum_i (\ln\omega_{p,i} - \omega_{p,i} e_{p,i} + 1)\Big)
- \mu P_{\mathrm{total}},
\label{eq:wmmse_objective}
\end{equation}
subject to $\mathcal{X}$ and the epigraph constraints $\bar{r}_c \le
\ln(1+\gamma_{c,i})$, which are also replaced by their WMMSE equivalents.

For fixed $\mathbf{u},\omega,\boldsymbol{\Phi},\mathbf{x}_{\mathrm{UAV}}$, the
resulting WMMSE objective is handled through block-coordinate updates of the
beamforming and power variables. The unit-norm conditions on the beamforming
vectors are explicitly enforced in the corresponding updates. For fixed
precoders and power allocation, the optimal receivers and weights are obtained
from closed-form MMSE expressions.

\subsection{Alternating Optimization and Convexification of BD-RIS Blocks}

We adopt an AO scheme inside each Dinkelbach iteration:

\begin{enumerate}
\item \textbf{Update receivers and weights:} With current precoders and RIS, compute the MMSE
equalizers $u_{c,i}, u_{p,i}$ and optimal weights $\omega_{c,i}, \omega_{p,i}$ via closed-form formulas.

\item \textbf{Update RSMA precoders and power:} With the receivers and weights
fixed, solve the resulting beamforming and power-allocation subproblem obtained
from~\eqref{eq:wmmse_objective}. The satellite power-allocation constraints and
the unit-norm beamforming conditions are explicitly enforced during the
corresponding block updates.

\item \textbf{Update BD-ARIS blocks:} Each group matrix $\boldsymbol{\Phi}_g$ appears
in quadratic terms $| \mathbf{g}_i^H \boldsymbol{\Phi} \mathbf{H}_u \mathbf{w}_k |^2$.
We linearize these terms around the current point using a first-order Taylor expansion, leading
to a convex approximation of the WMMSE objective with respect to $\boldsymbol{\Phi}_g$.
The per-group and total RIS power constraints~\eqref{eq:c4_active_group}--\eqref{eq:c4_active_total}
are convex (quadratic/SOC-representable) and are enforced exactly. A typical surrogate for
$q(\boldsymbol{\varphi})=\boldsymbol{\varphi}^T\mathbf{A}\boldsymbol{\varphi}$ is
$q(\boldsymbol{\varphi}) \approx q(\boldsymbol{\varphi}^{(t)}) + 2(\mathbf{A}\boldsymbol{\varphi}^{(t)})^T(\boldsymbol{\varphi}-\boldsymbol{\varphi}^{(t)})$.
Slack variables with an increasing quadratic penalty are used to handle temporary infeasibility.

\item \textbf{Update UAV horizontal position:} With all communication variables fixed, the
2-D variables $(x_{\mathrm{UAV}},y_{\mathrm{UAV}})$ are optimized via few iterations of SQP
(e.g., using \texttt{fmincon}). Analytical gradients are computed by chain rule through the
steering vectors; if unavailable, central finite differences are used. Warm-starting from the
previous position accelerates convergence.
\end{enumerate}

The AO loop runs until the change of the WMMSE objective falls below a threshold $\epsilon_{\mathrm{AO}}$.

\subsection{Algorithm Summary and Complexity}

The solver nests Dinkelbach (outer) and AO (inner) loops. Pseudocode is given in
Algorithm~\ref{alg:classical}. \textcolor{black}{Under standard SCA assumptions, the inner AO/SCA/SQP iterations converge to a stationary solution of the approximated subtractive subproblem, while Dinkelbach's outer iteration updates the fractional-programming parameter until the stopping criterion is satisfied. Since the overall EE maximization problem is non-convex, the resulting solution is not guaranteed to be the global optimum and is therefore used as an offline model-based stationary benchmark}~\cite{dinkelbach1967nonlinear}.

\begin{algorithm}[!t]
\caption{Dinkelbach + AO (WMMSE/SCA) + SQP Algorithm}
\label{alg:classical}
\begin{algorithmic}[1]
\STATE Initialize $\mathbf{w}_c,\{\mathbf{w}_i\},\boldsymbol{\Phi},a_c,\{a_i\},(x_{\mathrm{UAV}},y_{\mathrm{UAV}})$, $\mu=0$.
\REPEAT
  \REPEAT
    \STATE Update MMSE receivers $u$ and weights $\omega$ (closed-form).
\STATE Solve the beamforming and power-allocation subproblem for the RSMA precoders and power coefficients.
    \STATE Convexify $\boldsymbol{\Phi}$-dependent terms via SCA; update $\boldsymbol{\Phi}$.
    \STATE Update UAV position $(x_{\mathrm{UAV}},y_{\mathrm{UAV}})$ via SQP (2-variable smooth subproblem).
  \UNTIL{AO convergence}
  \STATE Update $\mu \leftarrow f/g$.
\UNTIL{$|f - \mu g| < \epsilon_{\mathrm{D}}$}
\end{algorithmic}
\end{algorithm}

\paragraph{Complexity \& implementation} Dominant cost: beamforming and power-allocation subproblems, with dense
linear-algebra operations scaling as approximately $\mathcal{O}(N^3)$ per
AO iteration; RIS updates scale with $G=M/2$ and are
highly parallelizable; SQP cost is negligible (2 variables). The convexified subproblems are solved via CVX/MOSEK. Tolerances: $\epsilon_{\mathrm{D}}=10^{-4}$,
$\epsilon_{\mathrm{AO}}\in[10^{-3},10^{-4}]$. \textcolor{black}{This algorithm provides high-quality stationary solutions that serve as an offline model-based benchmark.}

Even the Dinkelbach-AO solver requires more than $100$\,ms per instance, far
exceeding the LEO channel coherence time, and degrades under imperfect CSI,
motivating the learning-based approach for real-time, robust optimization.

\section{Deep Reinforcement Learning Approaches}
\label{sec:drl_solutions}

While the previous section presented a model-based solver, its iterative nature may limit
real-time applicability under fast-varying channels. To address this, we also develop
learning-based approaches that can quickly produce high-quality actions. We first formulate
the optimization problem as a Markov Decision Process (MDP) and then describe two DRL methods:
a standard TRPO baseline and a novel {Dinkelbach-Guided TRPO} that combines the strengths
of model-based optimization and policy gradient learning.

\subsection{MDP Formulation}
\label{sec:mdp_formulation}

The sequential decision process is modeled as a 4-tuple $(\mathcal{S},\mathcal{A},P,r)$,
where the agent (controller at the satellite/gateway) observes the current state,
selects an action, and receives a reward.

\subsubsection{State}
The state $\mathbf{s}_t \in \mathcal{S}$ captures the necessary channel information for
decision making. At time step $t$, the agent observes the estimated CSI of all links:
\begin{equation}
\mathbf{s}_t = \big\{ \widehat{\mathbf{h}}_1(t),\dots,\widehat{\mathbf{h}}_I(t),\;
\widehat{\mathbf{H}}_u(t),\;
\widehat{\mathbf{g}}_1(t),\dots,\widehat{\mathbf{g}}_I(t) \big\},
\end{equation}
where $\widehat{\mathbf{h}}_i$, $\widehat{\mathbf{H}}_u$, $\widehat{\mathbf{g}}_i$ denote
the imperfect channel estimates as described in Section~II.
When LEO satellite motion is considered, the state also implicitly captures the time evolution
of these fading coefficients.

\subsubsection{Action}
The action $\mathbf{a}_t \in \mathcal{A}$ comprises all optimization variables of
problem~\eqref{eq:opt_problem_RSMA}:
\begin{equation}
\mathbf{a}_t = \big\{ a_c,\, \{a_i\},\; \mathbf{w}_c,\, \{\mathbf{w}_i\},\;
\boldsymbol{\Phi},\; x_{\mathrm{UAV}},\, y_{\mathrm{UAV}} \big\}.
\end{equation}
Because many of these variables are complex-valued or subject to structural constraints,
we employ a {structured output layer} in the actor network:
\begin{itemize}
\item The power coefficients $(a_c, a_1,\dots,a_I)$ are obtained from a softmax output,
      automatically satisfying $a_c + \sum_i a_i = 1$ and $0\le a_c,a_i\le 1$.
\item Each beamforming vector $\mathbf{w}$ is normalized to unit norm after being generated
      as a real-plus-imaginary vector, ensuring $\|\mathbf{w}\|^2=1$; the actual transmit
      power is then controlled by $P_s a$ as in~\eqref{eq:transmitted_signal}.
\item The BD-RIS matrix $\boldsymbol{\Phi}_g$ for group $g$ is parameterized by its
      independent real-valued parameters: the log-amplitudes $\ln\beta_{g,1},\ln\beta_{g,2}$,
      phases $\theta_{g,1},\theta_{g,2}$, and the real and imaginary parts of $b_g$.
      The symmetric block is then constructed, and per-group power limits are enforced
      by clipping the gains if necessary.
\item The UAV position $(x_{\mathrm{UAV}},y_{\mathrm{UAV}})$ is directly output and then
      clamped to the feasible region~\eqref{eq:c9_active}--\eqref{eq:c10_active}.
\end{itemize}
This parameterization guarantees that all structural constraints are satisfied by design,
removing the need for penalty terms for these hard constraints during training.

\subsubsection{Reward}
The reward is the instantaneous EE, penalized only for the remaining
SINR constraints that are not automatically enforced:
\begin{equation}
{r_t} = \frac{{{\text{EE}}({{\mathbf{s}}_t},{{\mathbf{a}}_t})}}{{1 + \lambda \sum\limits_{i = 1}^I ( \max \{ 0,{\mkern 1mu} {\gamma_{\min}^{(c)}} - {\gamma _{c,i}}\}  + \max \{ 0,{\mkern 1mu} {\gamma_{\min}^{(i)}} - {\gamma _{p,i}}\} )}},
\label{eq:reward}
\end{equation}
where $\lambda$ is a penalty coefficient. When all SINR constraints are met, the denominator equals 1 and the reward equals the EE.

\textcolor{black}{In all experiments, the penalty coefficient is fixed at $\lambda=1$ and is not tuned per instance. This value is used identically for both Standard TRPO and DG-TRPO, allowing a direct comparison under the same reward. Most constraints (power coefficients, unit-norm precoders, UAV bounds) are enforced by construction through the actor network; only the SINR constraints are handled by the penalty term. We emphasize that SINR feasibility is primarily ensured by the supervised pretraining stage, where the actor is trained on Dinkelbach-AO solutions that are feasible by construction. The penalty term therefore serves mainly to maintain feasibility during fine-tuning, rather than to create it from scratch.}

\subsection{Standard TRPO Baseline}
\label{sec:trpo_standard}

Trust Region Policy Optimization (TRPO) is a state-of-the-art policy gradient algorithm
that guarantees monotonic improvement by constraining the Kullback-Leibler (KL) divergence
between successive policies~\cite{schulman2015trust}. For our MDP, we train a stochastic Gaussian policy
$\pi_{\boldsymbol{\theta}}(\mathbf{a}|\mathbf{s})$ whose mean is given by the structured
actor network described above and whose variance is a trainable diagonal matrix.

At each iteration, TRPO solves the constrained optimization problem
\begin{equation}
\begin{aligned}
\underset{\boldsymbol{\theta}}{\text{max}} \quad &
\mathbb{E}_{\mathbf{s},\mathbf{a}\sim\pi_{\boldsymbol{\theta}_{\text{old}}}}\!\left[
\frac{\pi_{\boldsymbol{\theta}}(\mathbf{a}|\mathbf{s})}
     {\pi_{\boldsymbol{\theta}_{\text{old}}}(\mathbf{a}|\mathbf{s})}
\hat{\mathcal{A}}(\mathbf{s},\mathbf{a}) \right] \\
\text{s.t.} \quad &
\mathbb{E}_{\mathbf{s}\sim\pi_{\boldsymbol{\theta}_{\text{old}}}}\!\left[
D_{\mathrm{KL}}\!\left(\pi_{\boldsymbol{\theta}_{\text{old}}}(\cdot|\mathbf{s})\,\|\,
                        \pi_{\boldsymbol{\theta}}(\cdot|\mathbf{s})\right) \right] \le \delta,
\end{aligned}
\end{equation}
where $\hat{\mathcal{A}}$ is the advantage estimate computed using a learned value function
$V_{\boldsymbol{\phi}}(\mathbf{s})$, and $\delta$ is the trust-region radius. The policy
update is performed via conjugate gradient and line search, ensuring the KL constraint
is respected. The value function is updated by regression on the empirical returns.

The standard TRPO agent learns purely from interaction with the environment, starting
from a randomly initialized policy. In our simulations, it serves as a pure learning-based
benchmark to assess how far a model-free approach can reach in this challenging
high-dimensional constrained problem.

\subsection{Proposed Dinkelbach-Guided TRPO (Hybrid Method)}
\label{sec:hybrid_trpo}

While standard TRPO can eventually discover effective policies, its training requires
extensive exploration and often converges to suboptimal local minima due to the problem's
non-convexity and tight constraints. To overcome these limitations, we propose
{Dinkelbach-Guided TRPO} (DG-TRPO), a hybrid optimization-learning scheme that
bootstraps the policy with high-quality solutions from the Dinkelbach solver and then
refines it via TRPO for robustness and real-time speed.

\subsubsection{Phase 1 – Supervised Pretraining with Dinkelbach Solutions}

\textcolor{black}{We first generate a pretraining dataset of $K=5000$ independent channel realizations by jointly randomizing the user geometry, UAV location, and channel fading conditions. Each ground user is independently and uniformly distributed over the considered disk of radius $50$\,km centered at the sub-satellite point, while the UAV horizontal coordinates are independently and uniformly sampled over the feasible region $[0,x_{\max}]\times[0,y_{\max}]$, with $x_{\max}=y_{\max}=5$\,km. For each realization, the NLoS components are generated independently according to the adopted CSCG model, and the Rician $K$-factors of the satellite--user and satellite--UAV links are uniformly sampled over $[4,12]$\,dB. Thus, the pretraining dataset spans the considered user and UAV geometries as well as a range of LoS/NLoS propagation conditions, rather than repeatedly using a single nominal channel configuration.}
 For each realization $\mathbf{s}_k$, the Dinkelbach+WMMSE/SCA+SQP algorithm of Section~IV is executed to
obtain a locally optimal action $\mathbf{a}_k^{\star}$ (the solution of
problem~\eqref{eq:opt_problem_RSMA} under perfect CSI). The actor network
$\pi_{\boldsymbol{\theta}}$ is then pretrained by minimizing the supervised loss
\begin{equation}
\mathcal{L}_{\mathrm{pre}}(\boldsymbol{\theta}) =
\frac{1}{K}\sum_{k=1}^{K} \big\| \boldsymbol{\mu}_{\boldsymbol{\theta}}(\mathbf{s}_k)
                               - \mathbf{a}_k^{\star} \big\|^2,
\label{eq:pretrain_loss}
\end{equation}
where $\boldsymbol{\mu}_{\boldsymbol{\theta}}(\mathbf{s})$ is the deterministic mean output
of the actor network (the stochastic policy's mean). This step initializes the policy
in the vicinity of high-EE solutions, dramatically reducing the exploration burden.

\textcolor{black}{It is important to note that the supervised pretraining stage is used as a warm-start rather than as the final policy-optimization stage. The subsequent TRPO fine-tuning adapts the pretrained actor through direct interaction with the noisy and time-varying environment. Therefore, the pretraining dataset primarily determines the quality of the initial policy and the exploration burden, while the fine-tuning stage provides further adaptation to conditions that may not be fully represented in the model-based demonstrations.}

\subsubsection{Phase 2 – TRPO Fine-Tuning}

After pretraining, we switch to the standard TRPO procedure described in
Section~\ref{sec:trpo_standard}, but using the pretrained weights as the starting point.
During this phase, the agent interacts with the \emph{actual environment}, which may include
CSI uncertainty, LEO satellite motion, and other dynamic effects that were not fully captured
by the static Dinkelbach solver. The trust-region constraint of TRPO keeps the policy
close to the pretrained solution, preventing catastrophic forgetting, while the reward
signal encourages adaptation to non-idealities.

Because the initial policy already starts in a high-EE region, the fine-tuning phase
requires significantly fewer episodes than training from scratch. Moreover, the resulting
policy inherits both the high performance of model-based optimization and the robustness
and millisecond inference time of a neural network.

\subsubsection{Computational Complexity and Real-Time Suitability}

Training of DG-TRPO is performed offline. Once trained, the actor network requires only
a single forward pass to map CSI to the optimal action. On a standard GPU, this inference
takes less than $0.5$\,ms, which is compatible with the effective CSI update interval adopted for the post-compensation LEO channel. In contrast, each Dinkelbach solve takes tens to
hundreds of milliseconds, making DG-TRPO more suitable for real-time control under satellite mobility and rapidly changing channels. The complexity comparison is
further discussed in the numerical results section.

\textcolor{black}{The main additional computational cost of DG-TRPO is the offline generation of the supervised dataset. Since one Dinkelbach-AO solution is obtained for each training realization, the preparation time scales approximately as $K_{\mathrm{pre}}T_{\mathrm{DAO}}$. With $K_{\mathrm{pre}}=5000$ and the reported per-realization Dinkelbach-AO time exceeding $100$\,ms, the sequential dataset-generation time exceeds approximately $500$\,s (about $8.3$ minutes). This cost is incurred only once during offline preparation and does not contribute to the online decision latency. Parallel execution can further reduce the wall-clock preparation time. After training, DG-TRPO requires only a single actor forward pass, with an inference time below $0.5$\,ms.}

The complete pseudocode of the proposed DG-TRPO method is provided 
in Algorithm~\ref{alg:dgtrpo}, while standard TRPO 
follows the original implementation in~\cite{schulman2015trust}.
Table~\ref{tab:complexity} summarizes the key features of the three proposed schemes.
\begin{table}[!t]
\centering
\caption{Comparison of the Three Proposed Optimization Schemes.}
\label{tab:complexity}
\footnotesize
\setlength{\tabcolsep}{4pt}
\begin{tabular}{p{1.7cm} p{1.9cm} p{1.9cm} p{1.9cm}}
\toprule
\textbf{Feature} & \textbf{Dinkelbach-AO} & \textbf{Standard TRPO} & \textbf{DG-TRPO (Proposed)} \\
\midrule
Type & Model-based, iterative & Model-free, online & Hybrid (pre-train + fine-tune) \\
Inference time & $>100$\,ms & $<0.5$\,ms & $<0.5$\,ms \\
CSI robustness & Sensitive & Learned (noisy training) & Learned (robust) \\
Training data & None & Many episodes & Moderate \\
Perfect CSI perf. & Reference & Comparable & Comparable \\
Noisy/mobile perf. & Degrades & Moderate & Robust \\
\bottomrule
\end{tabular}
\end{table}

\begin{algorithm}[!t]
\caption{Dinkelbach-Guided TRPO (DG-TRPO)}
\label{alg:dgtrpo}
\begin{algorithmic}[1]
\STATE \textbf{Phase 1: Supervised Pretraining}
\STATE Generate dataset $\mathcal{D} = \{(\mathbf{s}_k, \mathbf{a}_k^{\star})\}_{k=1}^{K}$ 
       using the Dinkelbach-AO solver (Algorithm~\ref{alg:classical}).
\STATE Pretrain actor network $\boldsymbol{\mu}_{\boldsymbol{\theta}}$ by minimizing 
       $\mathcal{L}_{\mathrm{pre}}(\boldsymbol{\theta})$ in~\eqref{eq:pretrain_loss}.
\STATE \textbf{Phase 2: TRPO Fine-Tuning}
\STATE Initialize policy $\pi_{\boldsymbol{\theta}}$ with pretrained weights,
       value function $V_{\boldsymbol{\phi}}$ randomly.
\REPEAT
    \STATE Collect trajectories using $\pi_{\boldsymbol{\theta}}$ in the actual 
           (noisy/mobile) environment.
    \STATE Compute advantages $\hat{\mathcal{A}}_t$ and returns $\hat{R}_t$.
    \STATE Update $\boldsymbol{\theta}$ via TRPO constrained gradient step
           (conjugate gradient + line search).
    \STATE Update $\boldsymbol{\phi}$ by minimizing $\|V_{\boldsymbol{\phi}}(\mathbf{s}) - \hat{R}\|^2$.
\UNTIL{convergence}
\end{algorithmic}
\end{algorithm}

In the simulation section, the three principal schemes, namely Dinkelbach-AO,
standard TRPO, and the proposed Dinkelbach-Guided TRPO, are evaluated together
with TRPO with random pretraining as an ablation baseline, and their performance
is compared across various metrics.

\section{Simulation Results}

We evaluate the three proposed schemes, Dinkelbach-AO, standard TRPO, and
DG-TRPO on the RSMA-enabled LEO satellite system assisted by a UAV-mounted BD-ARIS.
$I$ users are uniformly distributed in a $50$\,km radius around the sub-satellite point.

The Rician channel model with UAV-position-dependent LoS/NLoS components
(Section~II) is used, and LEO satellite motion is incorporated via a circular orbit.
Parameters are listed in Table~\ref{tab:parameters}. Both perfect and imperfect CSI
scenarios are considered (additive Gaussian errors with variance $\sigma_{\mathbf{X}}^2$).

\textcolor{black}{The nominal simulation setting uses $K_{\mathbf X}=8$\,dB, whereas the pretraining dataset deliberately randomizes the Rician $K$-factors over $[4,12]$\,dB to improve coverage of channel conditions. The final evaluation is performed on an independent set of 500 test realizations that are not included in the pretraining dataset.}

\textcolor{black}{In the simulations, the CSI update interval is set to $T_{\mathrm{CSI}}=1$ ms, which is a conservative CSI refresh interval for the post-compensation effective channel and is not intended to represent a universal raw LEO channel coherence time.}

All learning-based
methods are trained offline; the actor network of DG-TRPO is first pretrained on $K=5000$
channel realizations using the Dinkelbach-AO solutions, then fine-tuned with TRPO in the noisy,
time-varying environment. Standard TRPO is trained from scratch under the same conditions.
After training, we evaluate each scheme over 500 independent test realizations. For direct method-to-method comparison, the three schemes are evaluated on the same test realizations and, for each realization, use the same underlying channel and CSI estimate. To highlight
the benefits of the proposed RSMA+BD-ARIS combination, we also benchmark against NOMA-based
transmission as well as conventional RIS architectures (active diagonal RIS and passive RIS),
all optimized under the same power and SINR constraints. \textcolor{black}{The Dinkelbach-AO solver is executed per realization to provide an offline model-based reference for comparison.}

\textcolor{black}{Across the simulation figures, the proposed DG-TRPO is compared with five baselines: the model-based Dinkelbach-AO benchmark, the pure model-free Standard TRPO baseline, the TRPO with random pretraining ablation, NOMA with BD-ARIS as an alternative multiple-access scheme, and RSMA with active diagonal or passive RIS as alternative RIS architectures.}

\begin{table}[!t]
\centering
\caption{Simulation Parameters}
\label{tab:parameters}
\begin{tabular}{lll}
\toprule
\textbf{Parameter} & \textbf{Symbol} & \textbf{Value} \\
\midrule
Speed of light (m/s) & $c$ & $3\times10^8$ \\
Carrier frequency (Hz) & $f_c$ & $3.5\times10^9$ \\
Path loss exponent & $\ell$ & 2 \\
Variance of channel estimation error & $\sigma_{\mathbf{X}}^2$ & $10^{-2}$ \\
Maximum antenna gain (dBi) & $G_{\max}$ & 6.6 \\
Effective channel gain (average) & $\mathbb{E}[|\mathbf{H}_{\mathrm{eq},i}|^2]$ & $10^{-4}$ \\
AWGN noise power (W) & $\sigma^2$ & $10^{-10}$ \\
Rician $K$-factor (dB) & $K_{\mathbf{X}}$ & 8 \\
Penalty coefficient & $\lambda$ & 1 \\
Number of users & $I$ & 3 \\
Maximum satellite transmit power (dBm) & $P_{\text{SAT}}^{\max}$ & 56 \\
Satellite height (km) & $h_{\text{sat}}$ & 520 \\
Number of satellite antennas & $N$ & 32 \\
Circuit power (dBm) & $P_c$ & -10 \\
DC power (dBm) & $P_{DC}$ & -5 \\
Orbital inclination & $i_{\mathrm{orb}}$ & $53^\circ$ \\
CSI update interval (ms) & $T_{\mathrm{CSI}}$ & 1 \\
Minimum SINR (private messages) & $\gamma_{\min}^{(i)}$ & 0.5 \\
Minimum SINR (common message) & $\gamma_{\min}^{(c)}$ & 0.5 \\
Common-message power allocation (initial) & $a_c$ & 0.3 \\
Private-message power allocation (initial) & $a_i$ & 0.35 \\
Number of RIS elements & $M$ & 64 \\
Maximum RIS output power (dBm) & $P_{\mathrm{RIS}}^{\max}$ & 33 \\
Group-power scaling factor & $\kappa$ & 2 \\
RIS inverse amplifier efficiency & $\vartheta_{\mathrm{RIS}}$ & 1.25 \\
UAV height (km) & $h_{\text{UAV}}$ & 1 \\
Maximum UAV placement range in x & $x_{\max}$ & 5 km \\
Maximum UAV placement range in y & $y_{\max}$ & 5 km \\
UAV weight (N) & $W$ & 2 \\
Induced power correction factor & $k$ & 0.1 \\
Air density (kg/m$^3$) & $\rho$ & 1.225 \\
Profile drag coefficient &  $\delta$ & 0.02 \\
Rotor solidity & $s$ & 0.05 \\
Rotor disk area (m$^2$) & $A$ & 0.503 \\
Blade angular velocity (rad/s) & $\Omega$ & 300 \\
Rotor radius (m) & $R$ & 0.4 \\
System bandwidth (MHz) & $B$ & 5 \\
\bottomrule
\end{tabular}
\end{table}

\subsection{Convergence Performance}
\label{subsec:convergence}

\begin{figure}[!t]
    \centering
    \includegraphics[width=.8\columnwidth]{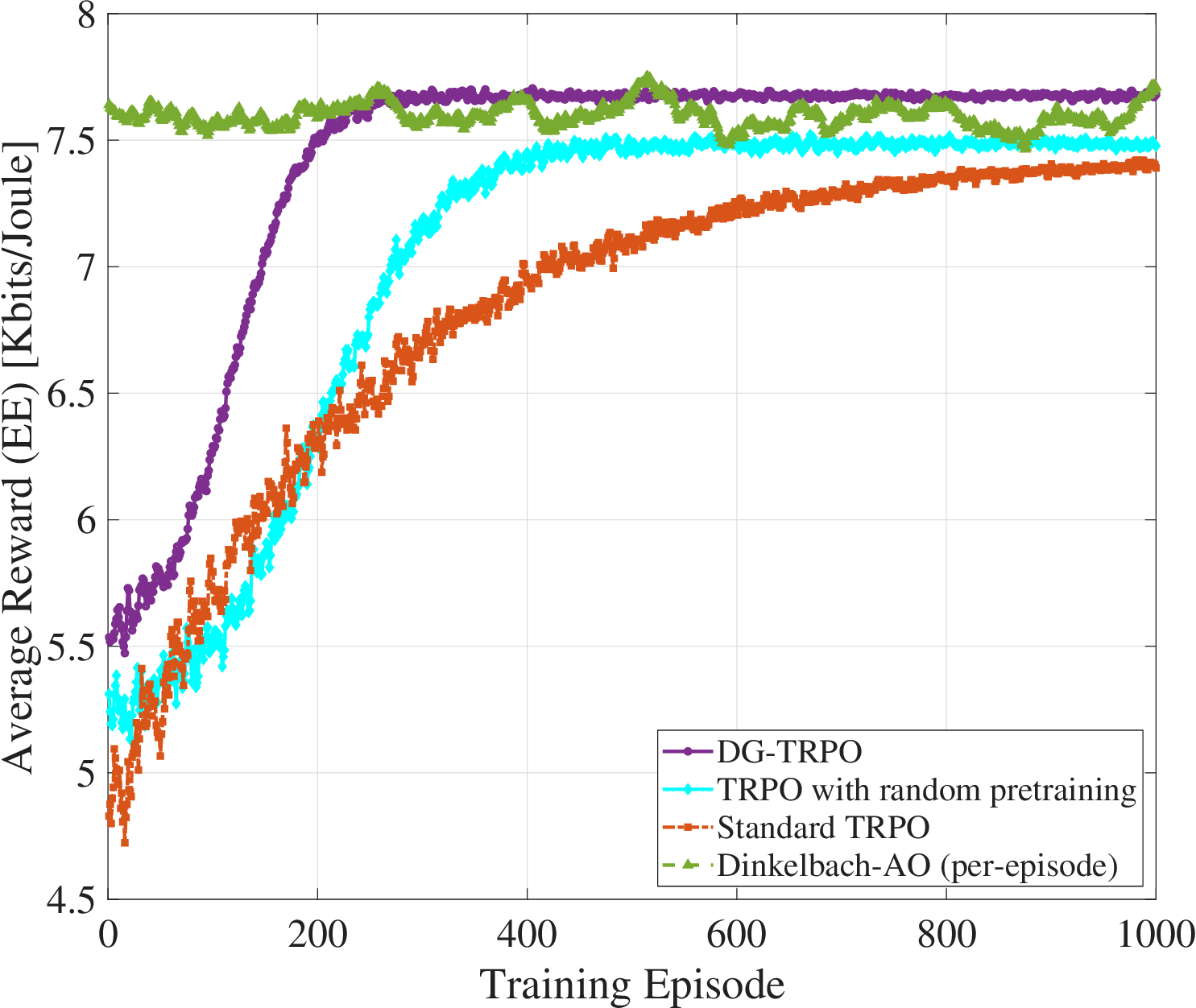}
    \caption{Convergence of DG‑TRPO, Standard TRPO, TRPO with random pretraining, and the Dinkelbach‑AO.}
    \label{fig:Rewards}
\end{figure}

Fig.~\ref{fig:Rewards} shows the learning curves of the three learning-based methods (DG-TRPO, Standard TRPO, and TRPO with random pretraining) alongside the per-episode EE obtained by the Dinkelbach-AO solver. The Dinkelbach‑AO curve exhibits moderate fluctuations
around its mean of $7.6$\,Kbits/Joule, reflecting natural variability of
channel realizations. \textcolor{black}{This model-based benchmark provides a reference for the reported EE comparison.}

\textcolor{black}{To isolate the effect of the Dinkelbach-guided pretraining, we introduce an intermediate baseline, TRPO with random pretraining, which uses the same actor architecture, the same pretraining dataset size, and the same TRPO fine-tuning as DG-TRPO, but with randomly generated feasible actions as pretraining targets instead of Dinkelbach-AO solutions.}

DG‑TRPO, initialized with supervised pretraining, starts at a reward of
approximately $5.5$\,Kbits/Joule and rises rapidly within the first $200$ episodes,
eventually stabilizing at about $7.68$\,Kbits/Joule, i.e., close to the Dinkelbach-AO average, with only a small difference in practical magnitude. The fast convergence and high asymptotic performance confirm that the pretrained policy is already situated in a favourable region of the parameter space and that the subsequent TRPO fine‑tuning effectively adapts it to the noisy, dynamic environment.

\textcolor{black}{The observation that DG-TRPO reaches an EE close to or slightly above the Dinkelbach-AO benchmark is consistent with the non-convex nature of the original problem. Dinkelbach handles the fractional objective, whereas the inner AO/WMMSE/SCA/SQP procedure provides a stationary solution without guaranteeing global optimality. Moreover, Dinkelbach-AO treats the available CSI estimate as exact during optimization, whereas DG-TRPO is trained to account for CSI uncertainty. Thus, a comparable or slightly higher EE for DG-TRPO in some regimes does not contradict the role of Dinkelbach-AO as a model-based stationary benchmark. To further examine the observed difference, a paired $t$-test was performed over the 500 test realizations, yielding a mean paired difference of $0.08$ Kbits/Joule with a 95\% confidence interval of $[0.058,\,0.102]$ Kbits/Joule ($t(499)=7.16$, $p<0.001$, $d_z=0.32$). The difference is therefore statistically detectable over the 500 realizations but remains small in practical magnitude.}

Standard TRPO, trained purely from scratch, begins at a much lower level (around
$4.9$\,Kbits/Joule) and requires approximately $800$ episodes to converge. Its
final reward settles near $7.45$\,Kbits/Joule, about $2\%$ below the Dinkelbach‑AO
mean. The slower convergence and lower asymptotic value illustrate the exploration
burden and the risk of sub‑optimal local minima in this high‑dimensional constrained
problem. Nevertheless, the curve is smooth and nearly flat after episode $700$,
indicating that the trust‑region constraint succeeds in stabilizing the learning
process.

\textcolor{black}{The intermediate baseline starts from a higher initial reward than Standard TRPO (approximately $5.3$\,Kbits/Joule), reflecting the structural benefit of pretraining: the actor has already learned to produce valid outputs. However, since its pretraining targets are random and therefore uninformative, it is occasionally surpassed by Standard TRPO during early training and converges only slightly above it (approximately $7.48$\,Kbits/Joule). This behavior is consistent with the known observation that supervised pretraining with uninformative targets can bias the policy and slow down exploration.}

\textcolor{black}{The comparison therefore separates the two effects. The gap between Standard TRPO and TRPO with random pretraining is small ($\approx 0.03$\,Kbits/Joule), showing that pretraining alone contributes marginally. In contrast, the gap between TRPO with random pretraining and DG-TRPO is substantial ($\approx 0.20$\,Kbits/Joule), confirming that the improvement comes from the Dinkelbach-guided content of the demonstrations, not from pretraining itself.}

{Overall, the convergence comparison shows that model-based pretraining places DG-TRPO in a high-EE region with substantially fewer training episodes, while subsequent TRPO fine-tuning enables adaptation to CSI imperfections and channel dynamics.}

\subsection{Energy Efficiency Analysis Under Varying Transmit Powers}

\begin{figure}[!t]
    \centering
    \begin{minipage}[b]{0.40\textwidth}
        \centering
        \includegraphics[width=\textwidth]{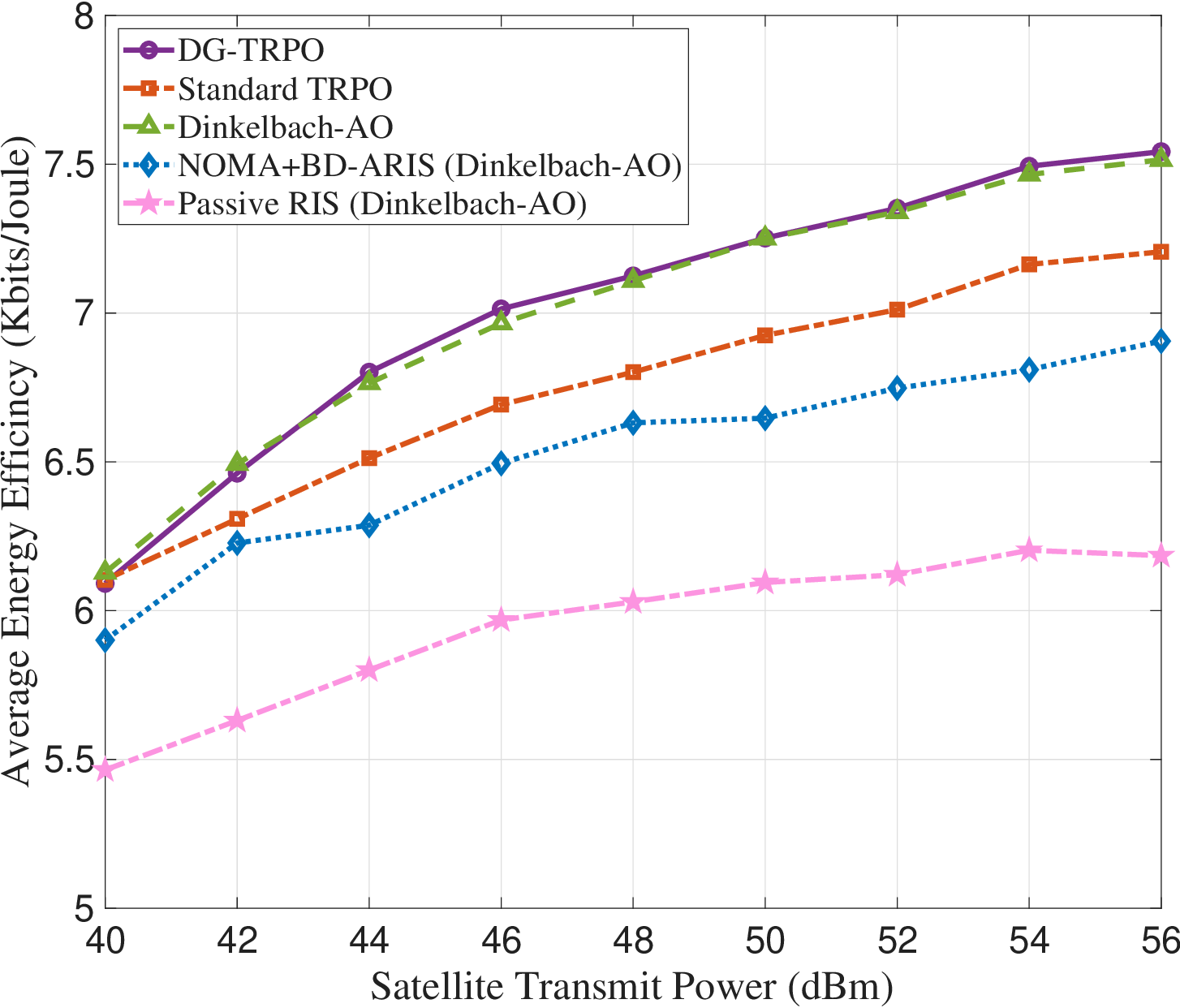}
        \vspace{0.5em}
        \textbf{(a)} EE vs.\ satellite transmit power
    \end{minipage}
    \hfill
    \begin{minipage}[b]{0.40\textwidth}
        \centering
        \includegraphics[width=\textwidth]{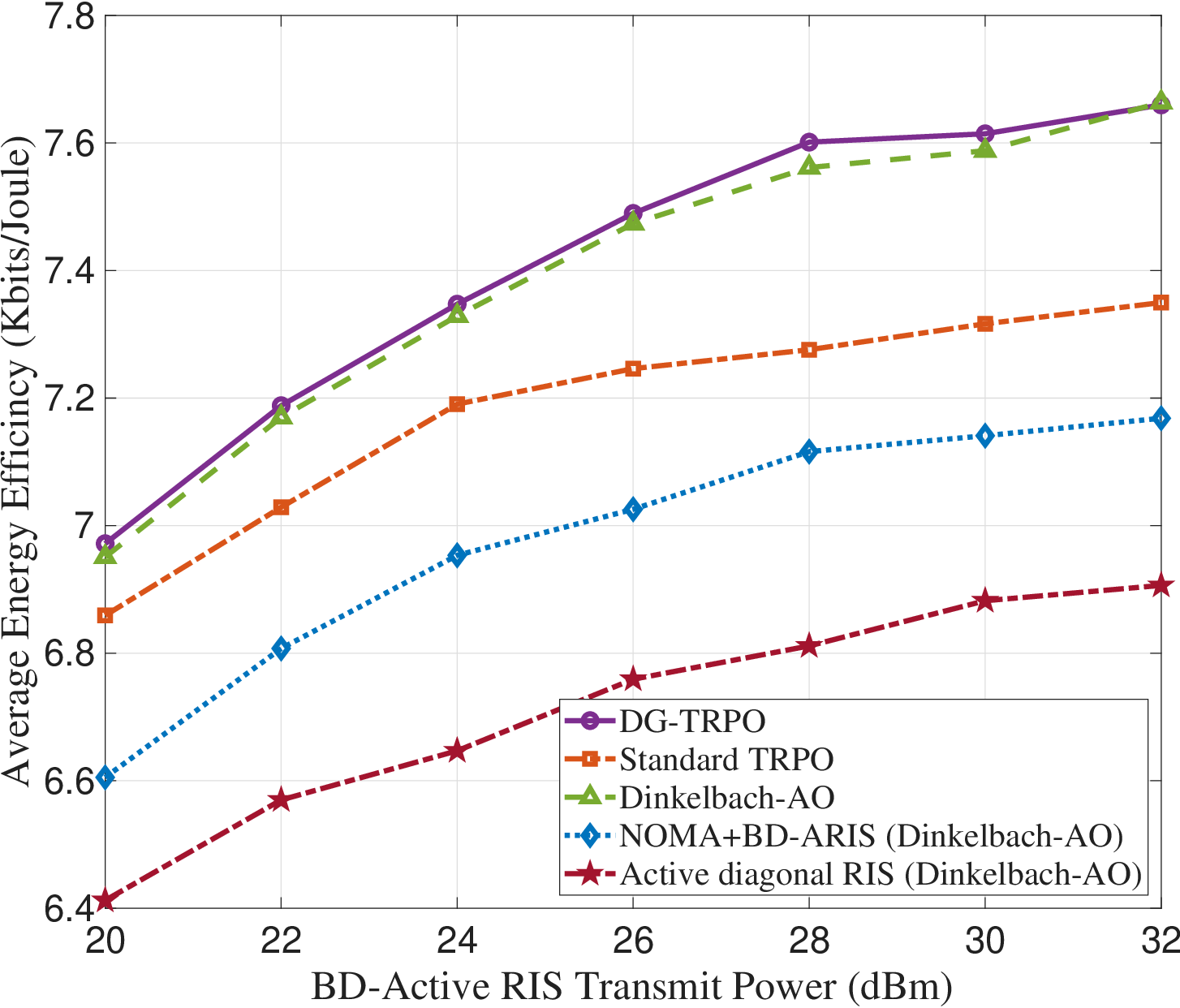}
        \vspace{0.5em}
        \textbf{(b)} EE vs.\ BD-ARIS transmit power
    \end{minipage}
\caption{EE comparison under varying transmit powers.
         (a) Satellite power sweep with passive RIS baseline;
         (b) BD‑ARIS power sweep with active diagonal RIS baseline.}
    \label{fig:ee_joint}
\end{figure}

Fig.~\ref{fig:ee_joint} compares the EE of the proposed DG‑TRPO, standard TRPO, and
Dinkelbach‑AO schemes against NOMA and conventional RIS baselines under sweeps of
satellite and BD‑ARIS transmit powers.

In Fig.~\ref{fig:ee_joint}(a), the satellite transmit power is increased from
$40$\,dBm to $56$\,dBm while the BD‑ARIS output power is fixed at $33$\,dBm.
All curves show a monotonic growth that gradually saturates as the SE
gains from higher transmit power are offset by the increased power consumption.
\textcolor{black}{DG-TRPO and the Dinkelbach-AO solver remain closely matched over the considered power range. This close agreement indicates that the hybrid pretraining places the policy in a high-EE region, while TRPO fine-tuning enables adaptation to non-ideal channel conditions.} 

Standard TRPO, which learns purely from interaction, reaches a slightly lower EE (by about $2\!-\!3\%$), reflecting the difficulty of navigating the high‑dimensional constrained optimization landscape without a warm start. The NOMA scheme with BD‑ARIS consistently underperforms all RSMA variants, with a gap that widens at higher power levels, illustrating the superior interference management capability of rate splitting.
The passive RIS baseline, which lacks signal amplification, attains the lowest EE
and exhibits a more gradual slope; its performance is primarily limited by the
weak direct satellite–user link and the absence of active power gain.

Fig.~\ref{fig:ee_joint}(b) varies the BD‑ARIS transmit power from $20$\,dBm to
$32$\,dBm, keeping the satellite power at $56$\,dBm. Because the satellite provides
a strong illuminating signal, even low RIS amplification yields a reasonably high
EE, and the curves saturate earlier than in the satellite‑power sweep. DG-TRPO again closely tracks the Dinkelbach-AO benchmark over the considered RIS power range, with differences that remain small in practical magnitude as quantified in section~\ref{subsec:convergence} . Standard TRPO stays marginally below DG‑TRPO,
and the NOMA scheme with BD‑ARIS suffers noticeable degradation, reaffirming the benefit of RSMA for multi‑user downlink. The active diagonal RIS (RSMA) baseline, which can amplify the incident wave but only with a conventional diagonal phase matrix, outperforms a purely passive surface yet remains clearly inferior to the group‑connected BD‑ARIS, highlighting the additional spatial degrees of freedom exploited by the beyond‑diagonal architecture.

\textcolor{black}{The preceding static evaluations are complemented by a mobility-aware simulation in which the satellite position, propagation distances, and satellite-related steering vectors are updated according to the time-varying LEO trajectory in \eqref{eq:satellite_trajectory}. At each CSI update interval $T_{\mathrm{CSI}}$, the nominal Doppler component is estimated and compensated, followed by pilot-assisted effective-CSI acquisition. The resulting estimated CSI is then provided to the DRL agents to generate the RSMA precoding, power allocation, and BD-ARIS configuration for the corresponding interval.}

\textcolor{black}{In contrast, the Dinkelbach-AO solver is re-executed whenever a new optimization point is required and therefore incurs substantially higher computational latency than the neural policies. Thus, the comparison under satellite mobility evaluates not only the static optimization quality but also the ability of each method to adapt its decisions to the evolving effective channel.}

\subsection{Impact of BD‑ARIS to User Distance on Sum Rate Performance}

\begin{figure}[!t]
    \centering
    \includegraphics[width=.8\columnwidth]{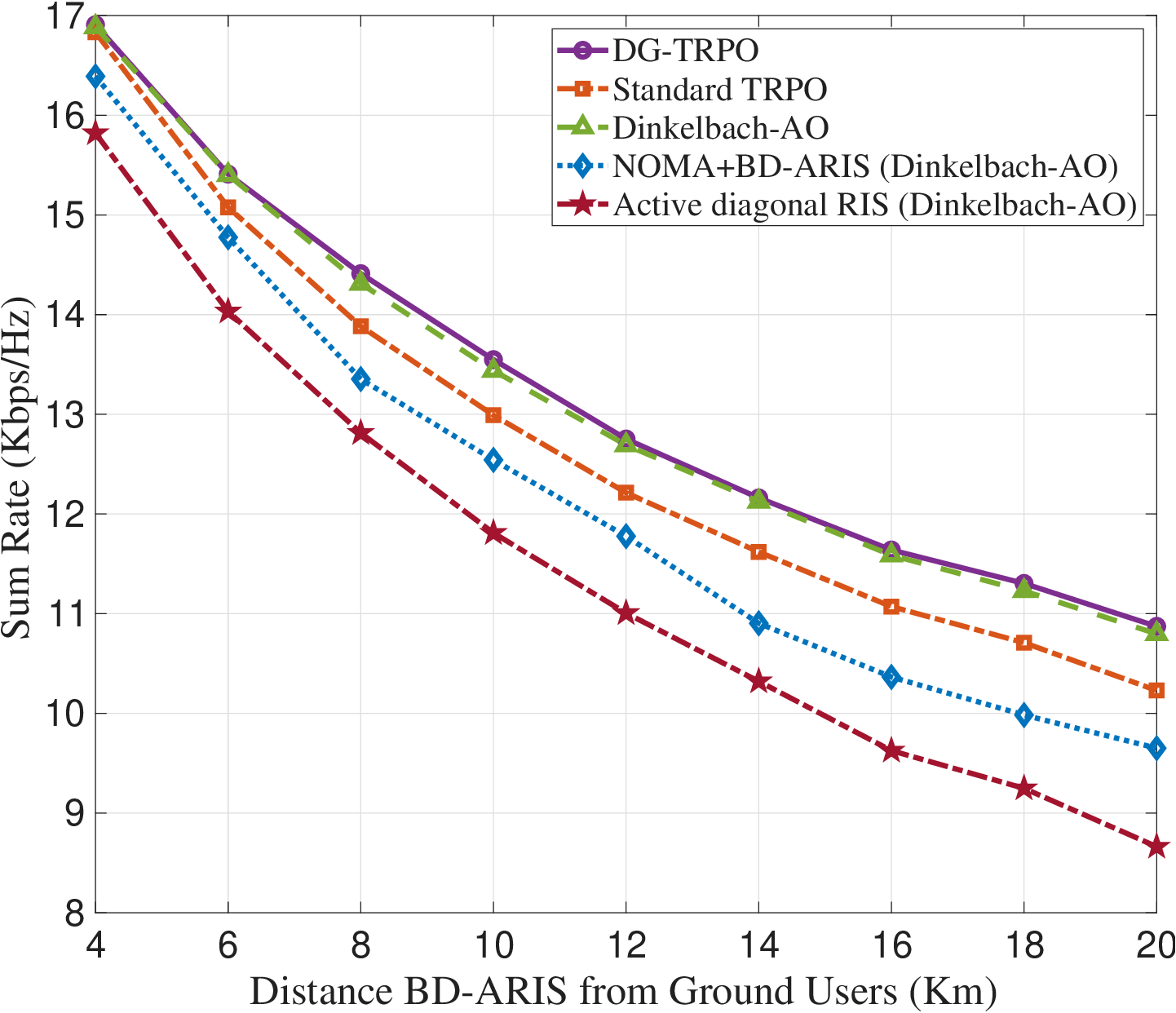}
    \caption{Sum rate versus distance between the UAV‑mounted BD‑ARIS and ground users.}
    \label{fig:SumRate_vs_Distance}
\end{figure}

Fig.~\ref{fig:SumRate_vs_Distance} presents the achievable sum rate versus the distance
between the UAV‑mounted BD‑ARIS and the ground users. The sum rate and energy efficiency are defined, respectively, as
\(
R_{\Sigma}
=
R_c+B\sum_{i=1}^{I}\log_2(1+\gamma_{p,i}) \)
and
\(
\mathrm{EE}=\frac{R_{\Sigma}}{P_{\mathrm{total}}} 
\).
 Five schemes are evaluated:
the proposed DG‑TRPO, standard TRPO, the Dinkelbach-AO solver serving as the model-based benchmark, and two baselines, namely NOMA combined with BD‑ARIS and RSMA with an active diagonal RIS, both optimized via Dinkelbach‑AO. The satellite transmit power is fixed at $56$\,dBm and the BD‑ARIS power at $33$\,dBm, while the distance is varied from $4$\,km to $20$\,km.

As the distance increases, all curves exhibit a monotonic decay due to the
growing path loss and reduced beamforming gain. \textcolor{black}{DG-TRPO closely tracks the Dinkelbach-AO benchmark across the entire range}. Standard TRPO suffers a moderate degradation, especially beyond
$10$\,km, highlighting the difficulty of learning optimal precoders and RIS
configurations from scratch without the model‑based warm start. The NOMA scheme
with BD‑ARIS falls further behind, as its SIC operation becomes less efficient
when the SNR imbalance among users is reduced at larger distances. The RSMA
system with an active diagonal RIS consistently underperforms the BD‑ARIS
variants, demonstrating that the group‑connected beyond‑diagonal architecture
preserves higher spatial multiplexing gains and is particularly beneficial when
the UAV–user links are weak. The divergence of the curves with distance confirms
that the combination of RSMA and BD‑ARIS, especially when driven by DG‑TRPO,
offers superior robustness against path‑loss‑induced performance degradation.

\subsection{Impact of CSI Error Variance on Communication Reliability}

\begin{figure}[!t]
    \centering
    \includegraphics[width=.87\columnwidth]{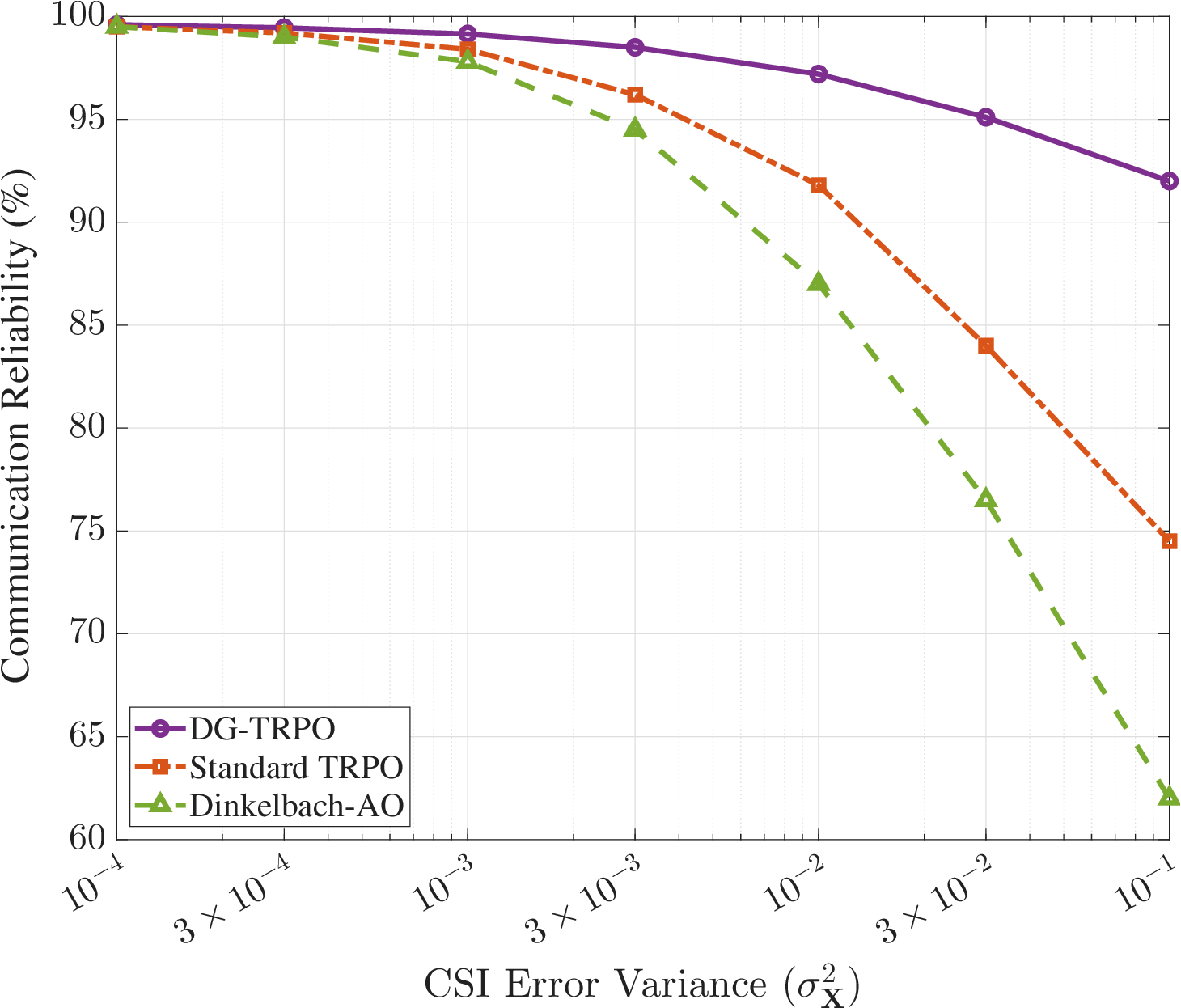}
    \caption{Communication reliability vs.\ CSI error variance for the three main schemes.}
    \label{fig:Reliability_CSI_Error_Variance}
\end{figure}

Fig.~\ref{fig:Reliability_CSI_Error_Variance} evaluates the communication reliability,
defined as the percentage of users that satisfy their minimum SINR requirements, as a
function of the CSI error variance $\sigma_{\mathbf{X}}^2$. A semilogarithmic scale
is used for the horizontal axis. Three schemes are compared: the proposed DG‑TRPO,
standard TRPO, and the Dinkelbach‑AO solver.

At $\sigma_{\mathbf{X}}^2 = 10^{-4}$ (near‑perfect CSI), all methods achieve a
reliability above $99\%$, confirming convergence to a high‑quality
solution when accurate channel knowledge is available. As $\sigma_{\mathbf{X}}^2$
increases, the reliability of all schemes monotonically degrades because
imperfect CSI leads to mismatched precoders, power allocations, and RIS
configurations, making it more likely that the target SINRs are not met.

The degradation rate, however, differs markedly among the methods.
Dinkelbach‑AO treats the estimated CSI as exact and solves a deterministic
optimization problem; therefore, its performance is highly sensitive to
estimation errors, and its reliability drops to $62\%$ at $\sigma_{\mathbf{X}}^2 = 10^{-1}$.
Standard TRPO, trained from scratch with noisy channel realisations, exhibits
moderate robustness, retaining $74.5\%$ reliability at the same error level.
DG‑TRPO is the most resilient scheme: its reliability remains above
$92\%$ even at $\sigma_{\mathbf{X}}^2 = 10^{-1}$, losing less than $8\%$
compared with the perfect‑CSI case. This superior robustness stems from the
two‑stage training procedure: the supervised pretraining on Dinkelbach‑AO
solutions provides a strong baseline policy, while the subsequent TRPO
fine‑tuning in a realistically noisy environment teaches the actor network to
produce decisions that remain effective despite CSI imperfections.

\textcolor{black}{Further confirmation of the role of pretraining comes from the fact that Standard TRPO and DG-TRPO use the same reward function with the same penalty coefficient $\lambda=1$, and differ only in the pretraining stage. Despite this identical $\lambda$, Standard TRPO reaches only $74.5\%$ reliability, whereas DG-TRPO reaches $92.0\%$. The gap of $17.5$ percentage points therefore cannot be attributed to $\lambda$; it is a direct consequence of the supervised pretraining on Dinkelbach-AO solutions. This confirms that the high reliability of DG-TRPO is a property of the training architecture, not a coincidence of the chosen $\lambda$.}

These results confirm that the DG‑TRPO agent not only accelerates convergence
and boosts nominal EE, but also enhances communication reliability in the
presence of channel estimation errors, which is a critical requirement for
practical LEO satellite systems.

\subsection{Scalability with Varying Number of Users}

\begin{figure}[t]
  \centering
    \includegraphics[width=.8\linewidth]{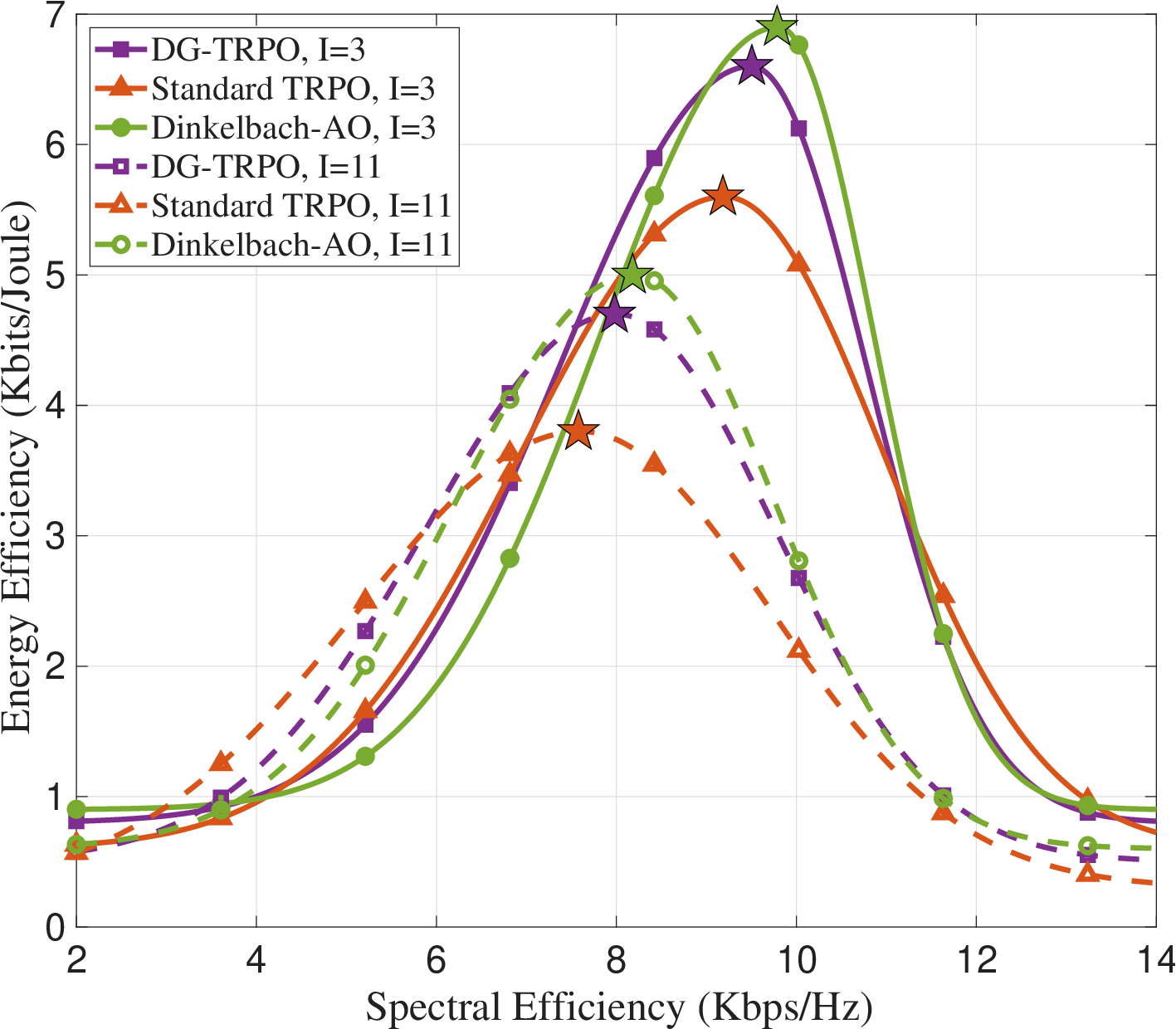}
  \caption{EE vs.\ SE for $I=3$ (solid) and $I=11$ (dashed) users.}
  \label{fig:scalability}
\end{figure}

Fig.~\ref{fig:scalability} shows the EE versus SE
for the three main schemes when the network load is varied. Solid lines correspond
to a lightly loaded scenario with $I=3$ users, while dashed lines represent a dense
configuration with $I=11$ users. All curves are obtained by sweeping the satellite
transmit power and evaluating the resulting EE–SE pairs.

The curves exhibit the characteristic peaked shape of the EE–SE trade‑off: as the
transmit power increases, both EE and SE first improve together, but after a certain
point the growth in power consumption overtakes the SE gain, causing
EE to decline. At $I=3$, the peaks are higher and occur at larger SE values because
the limited multi‑user interference allows the RSMA precoders and the BD‑ARIS to
focus energy more effectively. When the number of users grows to $I=11$, the
available spatial and power resources must be shared among more nodes, increasing
interference and reducing the achievable rate per unit power. Consequently, all
curves shift leftwards and downwards, and the peak EE drops noticeably.

Among the three methods, DG‑TRPO achieves the highest EE peak for both
user densities and retains the largest SE value at the peak, confirming its
superior capability to balance common and private rate allocations even in dense
scenarios. The standard TRPO baseline suffers a larger degradation, particularly
in the high‑SE region where its policy, learned from scratch, deviates further from the model-based benchmark. \textcolor{black}{The Dinkelbach-AO solver serves as a model-based benchmark, and the EE gap relative to DG-TRPO remains marginal at $I=3$ and almost vanishes at $I=11$, illustrating the limits of a static model-based approach under strong interference.} These results
demonstrate that the proposed DG‑TRPO framework scales gracefully with the user
population, making it a promising candidate for future IoT‑over‑satellite
deployments where the user density may vary significantly.

\subsection{Influence of RIS Architecture and Number of Elements}

\begin{figure}[!t]
    \centering
    \includegraphics[width=.8\columnwidth]{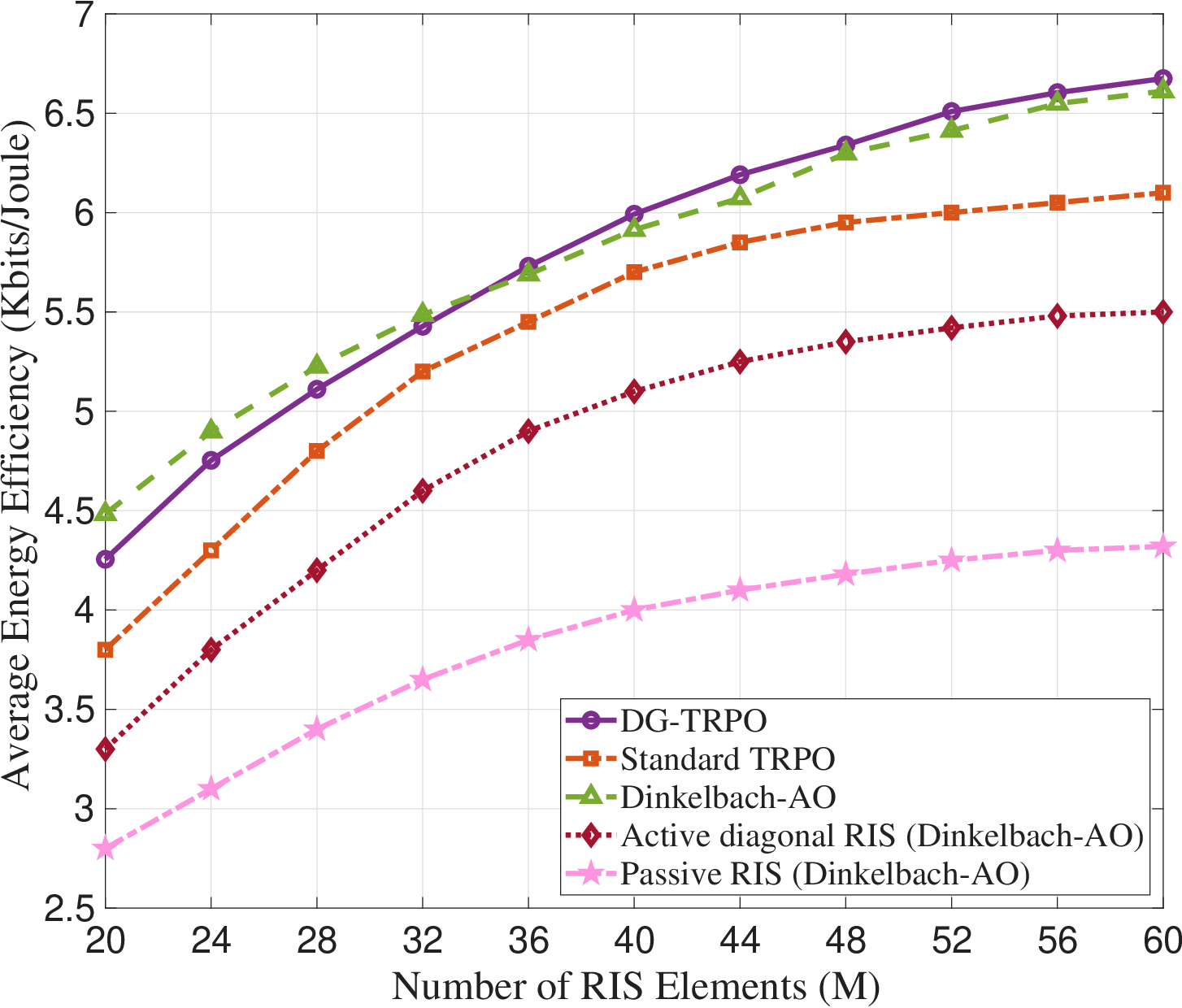}
    \caption{Energy efficiency versus number of RIS elements for different RIS architectures and optimization methods.}
    \label{fig:ee_vs_ris_size}
\end{figure}

Fig.~\ref{fig:ee_vs_ris_size} examines the EE when
RIS elements $M$ grows from $20$ to $60$. Five configurations are compared:
DG‑TRPO with BD‑ARIS, standard TRPO with BD‑ARIS, the Dinkelbach-AO solver with BD-ARIS (model-based benchmark), and two baselines employing RSMA
optimised via Dinkelbach‑AO: an active diagonal RIS and a passive RIS. The
satellite and BD‑ARIS transmit powers are fixed at $56$\,dBm and $33$\,dBm.
All curves show a monotonic increase with $M$, because a larger RIS provides
more spatial degrees of freedom and a higher aperture gain. The passive RIS
achieves only a marginal improvement, as it cannot compensate for the deep
path loss of the satellite link without amplification. The active diagonal RIS
benefits from signal amplification, but its EE saturates early, reaching only
$5.5$\,Kbits/Joule at $M=60$, because a conventional diagonal phase matrix
cannot exploit inter‑element coupling.

\textcolor{black}{It is important to note that the reported EE for all RIS architectures already accounts for the RIS internal power consumption through $P_{\mathrm{BD-ARIS}}$ in \eqref{eq:total_power}. The comparison in Fig.~\ref{fig:ee_vs_ris_size} therefore reflects a genuine net trade-off. The passive RIS, despite its nearly zero internal power consumption, attains the lowest EE because it cannot compensate for the deep path loss of the satellite link. The active diagonal RIS achieves a moderate EE improvement, but its circuit power and limited spatial flexibility restrict its performance. In contrast, the proposed group-connected BD-ARIS attains the highest EE because (i) the group-connected structure reduces the number of active amplifiers to $M/2$, lowering the static circuit power, and (ii) the amplification gain simultaneously increases the achievable rate and reduces the required satellite transmit power, more than offsetting the additional RIS power consumption.}

The three BD‑ARIS schemes substantially outperform the diagonal and passive
configurations. Among them, standard TRPO yields the lowest EE, despite using
the same BD‑ARIS hardware, because its policy, trained from scratch, struggles
to simultaneously configure the precoders, power allocation, and RIS matrix
when the action space grows with $M$. Dinkelbach-AO provides a model-based reference that rises quickly but saturates for $M \ge 48$, indicating that the model‑based
solver can no longer fully harness the additional elements.

\textcolor{black}{DG-TRPO, in contrast, starts slightly below Dinkelbach-AO at small $M$ and gradually approaches the model-based benchmark as $M$ increases}, reaching approximately $7.1$\,Kbits/Joule at $M=60$.
This behavior demonstrates that the fine-tuned neural policy effectively learns how to exploit the extra degrees of freedom of larger BD-ARIS panels, especially in the presence of coupling and hardware imperfections, whereas the deterministic SCA-based optimization of Dinkelbach-AO becomes conservative. This result confirms that the group-connected beyond-diagonal architecture, when controlled by the DG-TRPO agent, scales favourably with the RIS size and can unlock the full potential of large intelligent surfaces in satellite downlink scenarios.

\section{Conclusion}

This paper investigated an RSMA-enabled non-terrestrial communication system
composed of a LEO satellite, a UAV-mounted BD-ARIS, and multiple ground users.
To jointly optimize the RSMA precoders, power allocation, RIS configuration, and UAV
placement for energy efficiency maximization, we developed three solutions:
a model-based iterative algorithm relying on Dinkelbach's method, WMMSE,
SCA, and SQP; a standard TRPO baseline trained from scratch; and a novel
Dinkelbach‑Guided TRPO (DG‑TRPO) that initializes the policy via supervised
pretraining on model-based solutions and then fine-tunes it with TRPO.

\textcolor{black}{Simulation results showed that the model-based solver provides an offline stationary reference under its assumed CSI conditions, but its high computational delay ($>100$\,ms) and sensitivity to channel uncertainty hinder real-time LEO operation.} \textcolor{black}{Standard TRPO converges to a competitive EE, yet it requires extensive training and remains below the model-based reference in the considered settings.}
DG‑TRPO, by combining model‑based initialization with policy gradient fine‑tuning,
\textcolor{black}{achieves EE comparable to the model-based reference while providing rapid convergence and sub-millisecond inference,} making it suitable for
LEO satellite dynamics. It also demonstrated high reliability under CSI uncertainty, retaining
over $92\%$ reliability at the largest considered estimation variance, and
supports operation under the considered LEO satellite mobility. Furthermore, the
RSMA and BD‑ARIS combination consistently outperformed NOMA and conventional
RIS architectures, demonstrating superior interference management and spatial
multiplexing gains.

These findings establish the proposed DG‑TRPO algorithm, together with the RSMA
BD‑ARIS architecture, as a scalable, robust, and energy‑efficient solution for
future 6G massive IoT non‑terrestrial networks.

\section*{Acknowledgment}

This work is based upon research funded by the INSF, the R\&D Center of MCI, and the TDCCC, under Project No.~4030485.

The work of E. Jorswieck was supported partly by the Federal Ministry of Research, Technology and Space (BMFTR), Germany, through the Project xG-RIC under Grant 16KIS2435. The work of E. Jorswieck was carried out in part within the Nitride Technology Center's NTC BRIGHT initiative (Grant No. ZN4786), funded by zukunft.niedersachsen, the joint science funding program of the Lower Saxony Ministry of Science and Culture and the Volkswagen Foundation.

\bibliographystyle{IEEEtran}
\bibliography{StarRIS}

\begin{IEEEbiography}[{\includegraphics[width=1in,height=1.25in,clip,keepaspectratio]{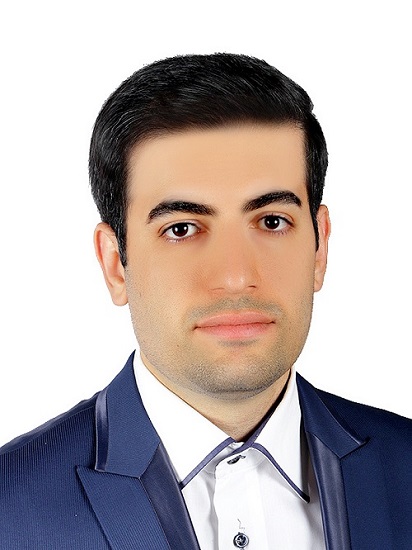}}]{Rahman Saadat Yeganeh}
received his Ph.D. degree in Electrical Engineering (Telecommunication Systems) from Isfahan University of Technology (IUT), Isfahan, Iran, in 2024. He subsequently conducted postdoctoral research in telecommunication systems at Sharif University of Technology (SUT), Tehran, Iran. His research and professional activities encompass the design, optimization, and performance analysis of wireless and satellite communication systems, as well as Internet of Things (IoT) networks. His research interests include next-generation wireless communications, with a particular focus on 6G networks, satellite communications, IoT, symbiotic radio, reconfigurable intelligent surfaces (RISs), non-orthogonal multiple access (NOMA), rate-splitting multiple access (RSMA), and signal processing. He is also interested in intelligent resource allocation and learning-based optimization techniques for advanced communication networks.
\end{IEEEbiography}

\begin{IEEEbiography}[{\includegraphics[width=1in,height=1.25in,clip,keepaspectratio]{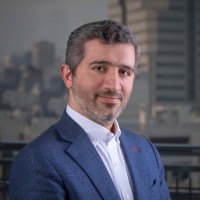}}]{Hamid Behroozi}
(Member, IEEE) received the B.Sc. degree in electrical engineering from the University of Tehran, Tehran, Iran, in 2000, the M.Sc. degree in electrical engineering from the Sharif University of Technology, Tehran, in 2003, and the Ph.D. degree in electrical engineering from Concordia University, Montreal, QC, Canada, in 2007. From 2007 to 2010, he was a Postdoctoral Fellow with the Department of Mathematics and Statistics, Queen’s University, Kingston, ON, Canada. He is currently an Associate Professor with the Department of Electrical Engineering, Sharif University of Technology. His research interests include information theory, joint source-channel coding, artificial intelligence in signal processing and data science, and cooperative communications. He was the recipient of several academic awards, including the Ontario Postdoctoral Fellowship awarded by the Ontario Ministry of Research and Innovation, the Quebec Doctoral Research Scholarship awarded by the Government of Quebec, the Hydro Quebec Graduate Award, and the Concordia University Graduate Fellowship
\end{IEEEbiography}

\begin{IEEEbiography}[{\includegraphics[width=1in,height=1.25in,clip,keepaspectratio]{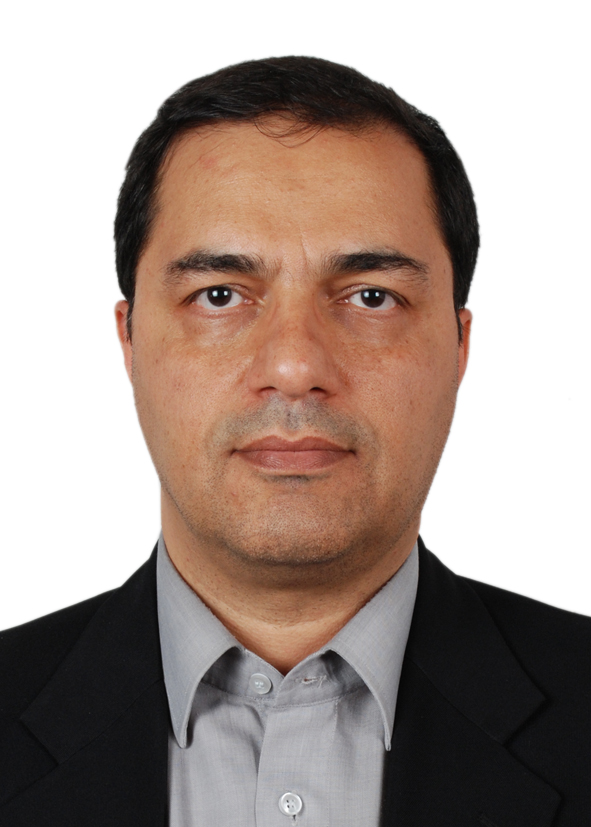}}]{Mohammad Javad Omidi}
received the Ph.D. degree from the University of Toronto in 1998. He has extensive industry experience in Canada, specializing in the design of broadband communication systems. He has held several prominent academic and administrative positions with the Isfahan University of Technology, Iran, including as a Professor with the Department of Electrical and Computer Engineering (ECE), the Chair of the IT Center, the Chair of the ECE Department, and the Vice President for Research and Technology.

He currently serves as a Professor with the ECE Department, Kuwait College of Science and Technology. Beyond academia, he has played a significant role in fostering innovation and entrepreneurship. He has extensive experience in managing science parks, supporting technology start-ups, and mentoring university graduates in their entrepreneurial ventures. His research focuses on wireless communications, digital communication systems, and cognitive radio systems. He has authored numerous publications and holds six U.S. patents along with four international patents in these areas.
\end{IEEEbiography}
\begin{IEEEbiography}[{\includegraphics[width=1in,height=1.25in,clip,keepaspectratio]{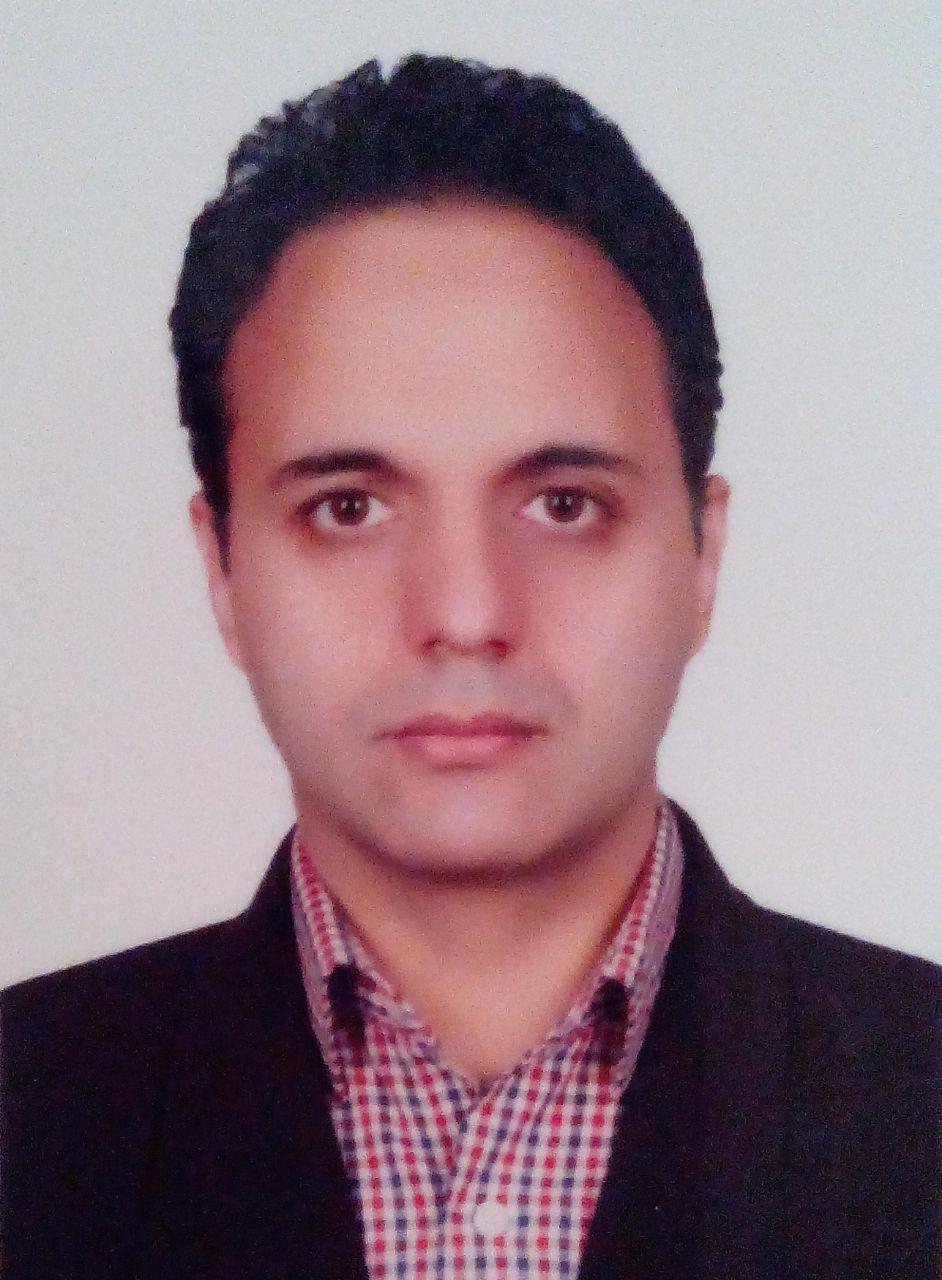}}]{Mohammad Robat Mili}
received the Ph.D. degree in electrical and electronic engineering from the University of Manchester, U.K., in 2012. He held postdoctoral research positions at the Department of Telecommunications and Information Processing, Ghent University, Belgium and the Department of Electrical Engineering, Sharif University of Technology, Iran. His main research interests are in the area of design and analysis of wireless communication networks with particular focus on 5G and 6G cellular networks using mathematical methods such as optimization theory, game theory, and machine learning
\end{IEEEbiography}

\begin{IEEEbiography}
[{\includegraphics[width=1in,height=1.25in,clip,keepaspectratio]{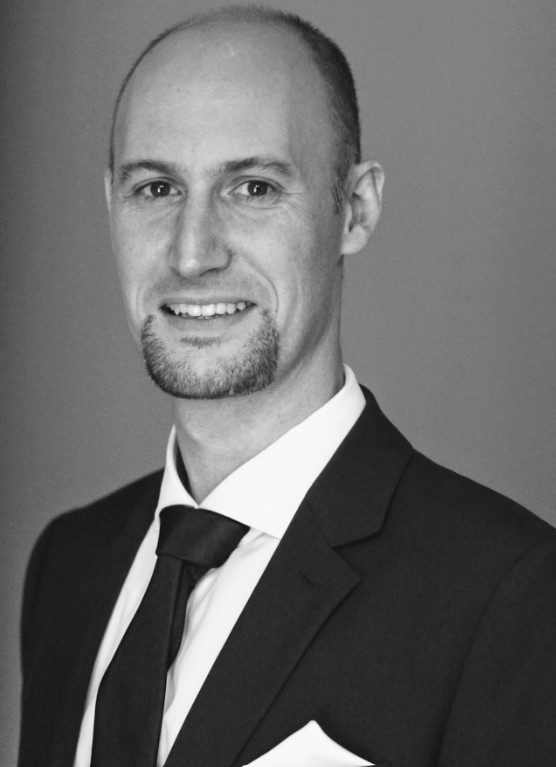}}]{Symeon Chatzinotas} 
is currently Full Professor / Chief Scientist I and Head of the research group SIGCOM in the Interdisciplinary Centre for Security, Reliability and Trust, University of Luxembourg. In parallel, he is an Adjunct Professor in the Department of Electronic Systems, Norwegian University of Science and Technology, an Eminent Scholar of the Kyung Hee University, Korea and a Collaborating Scholar of the Institute of Informatics \& Telecommunications, National Center for Scientific Research “Demokritos”. 

In the past, he has been a Visiting Professor at EPFL, Switzerland and University of Parma, Italy and contributed in numerous R\&D projects for the Institute of Telematics and Informatics, Center of Research and Technology Hellas and Mobile Communications Research Group, Center of Communication Systems Research, University of Surrey.

He has received the M.Eng. in Telecommunications from Aristotle University of Thessaloniki, Greece and the M.Sc. and Ph.D. in Electronic Engineering from University of Surrey, UK in 2003, 2006 and 2009 respectively. 

He has authored more than 800 technical papers in refereed international journals, conferences and scientific books and has received numerous awards and recognitions, including the IEEE Fellowship, IEEE Distinguished Contributions Award and IEEE Harry Rowe Mimno Award. He has served in the editorial board of npj Wireless Technology, IEEE Transactions on Communications, IEEE Open Journal of Vehicular Technology and the International Journal of Satellite Communications and Networking.
\end{IEEEbiography}

\begin{IEEEbiography}
[{\includegraphics[width=1in,height=1.25in,clip,keepaspectratio]{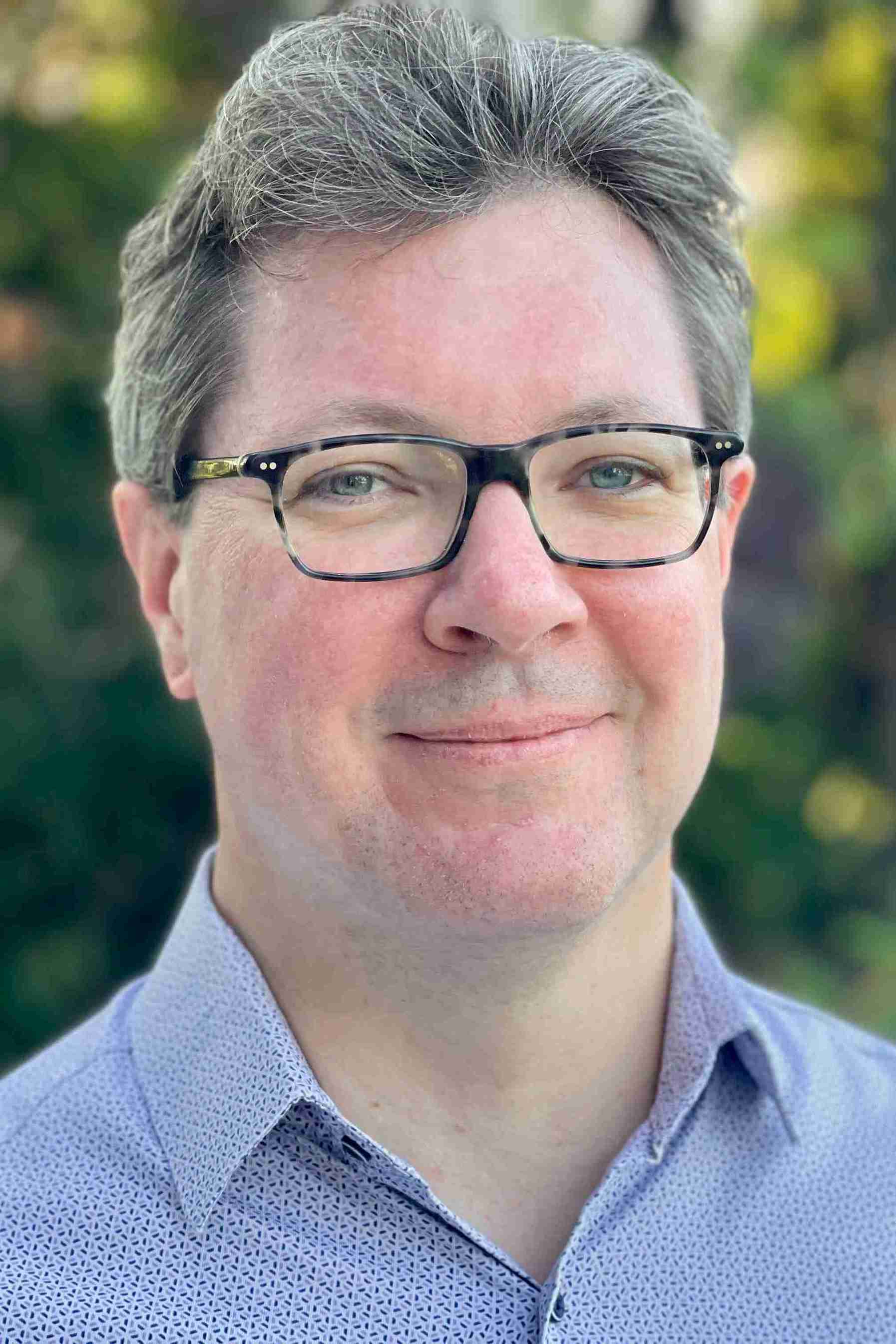}}]{ Eduard Axel Jorswieck}
is full professor with the Faculty of Electrical Engineering, Information Technology, Physics of TU Braunschweig. He is the Managing Director of the Institute for Communications Technology at TU Braunschweig, Germany. From 2008 until 2019, he was the head of the Chair for Communications Theory and Full Professor at TU Dresden, Germany. He is IEEE Fellow. His general interests are in signal processing for communications and networking, applied information theory and communication theory. He has published more than 220 journal articles, 20 book chapters, one book, four monographs, and some 350 conference papers. He was a co-recipient of the IEEE Signal Processing Society Best Paper Award 2006. His colleagues and he were also recipients of the Best Paper Awards and the Best Student Paper Awards from the IEEE CAMSAP 2011, IEEE WCSP 2012, IEEE SPAWC 2012, IEEE ICUFN 2018, PETS 2019, ISWCS 2019, and IEEE ICC 2024. 

Since 2024, he has been an Editor for IEEE TRANSACTIONS ON INFORMATION THEORY. He was on the editorial boards of the IEEE SIGNAL PROCESSING LETTERS, IEEE TRANSACTIONS ON SIGNAL PROCESSING, the IEEE TRANSACTIONS ON WIRELESS COMMUNICATIONS, IEEE TRANSACTIONS ON INFORMATION FORENSICS AND SECURITY, and IEEE TRANSACTIONS ON COMMUNICATIONS. From 2017 until 2026, he was the Editor-in-Chief of the Springer Journal on Wireless Communications and Networking.
\end{IEEEbiography}

\end{document}